\documentclass[fleqn,10pt]{wlscirep}
\usepackage[utf8]{inputenc}
\usepackage[T1]{fontenc}

\usepackage{pgfplots}
\usepgfplotslibrary{statistics}
\usepackage{filecontents}

\usepackage{multirow}
\usepackage{booktabs}
\usepackage{rotating}
\usepackage[normalem]{ulem}
\usepackage{tabularray}
\usepackage{makecell}
\usepackage{colortbl}

\usepackage{graphicx}%
\usepackage{multirow}%
\usepackage{amsmath,amssymb,amsfonts}%
\usepackage{amsthm}%
\usepackage{mathrsfs}%
\usepackage[title]{appendix}%
\usepackage{xcolor}%
\usepackage{textcomp}%
\usepackage{manyfoot}%
\usepackage{booktabs}%
\usepackage{algorithm}%
\usepackage{algorithmicx}%
\usepackage{algpseudocode}%
\usepackage{listings}%
\usepackage{subcaption}
\usepackage{adjustbox}
\usepackage{comment}
\usepackage{etoolbox}
\usepackage{ellipsis} 
\usepackage{xstring}
\usepackage{xparse} 

\newcommand{\rstwww}{\url{www.rockstartit.com}}
\newcommand{\rstname}{\emph{RockStartIT}}

\newcommand{\cellgrey}{{\cellcolor[rgb]{0.937,0.937,0.937}}}
\usepackage{pgfplots}
\usepgfplotslibrary{fillbetween}

\usepackage{xcolor}
\newcommand{\apos}{\textcolor{teal}{$\uparrow$}}
\newcommand{\appos}{\textcolor{teal}{$\uparrow\uparrow$}}
\newcommand{\adown}{\textcolor{orange}{$\downarrow$}}
\newcommand{\addown}{\textcolor{orange}{$\downarrow\downarrow$}}
\newcommand{\aupdown}{\textcolor{blue}{$\updownarrow$}}

\usepackage{caption}
\usepackage{subcaption}

\usepackage{lscape}
\usepackage{booktabs}
\usepackage{makecell}
\usepackage{rotating}
\usepackage{multirow}
\usepackage{colortbl}
\usepackage{vcell}
\usepackage{hhline}
\usepackage{graphicx}

\title{Measuring Computer Science Enthusiasm: A Questionnaire-Based Analysis of Age and Gender Effects on Students’ Interest}
\author[1,2]{Kai Marquardt}
\author[3]{Robert Hanak}
\author[1]{Anne Koziolek}
\author[1,2,3]{Lucia Happe}
\affil[1]{Karlsruhe Institute of Technology, Department of Informatics, Karlsruhe, 76131, Germany}
\affil[2]{Karlsruhe Institute of Technology, Department for Interdisciplinary Didactics, Karlsruhe, 76131, Germany}
\affil[3]{University of Economics, Faculty of Business Management, Bratislava, 81101, Slovakia}

\begin{abstract}
This study offers new insights into students’ interest in computer science (CS) education by disentangling the distinct effects of age and gender across a diverse adolescent sample. Grounded in the person–object theory of interest (POI), we conceptualize enthusiasm as a short-term, activating expression of interest that combines positive affect, perceived relevance, and intention to re-engage. Experiencing such enthusiasm can temporarily shift CS attitudes and strengthen future engagement intentions, making it a valuable lens for evaluating brief outreach activities.
To capture these dynamics, we developed a theoretically grounded questionnaire for pre–post assessment of the enthusiasm potential of CS interventions. Using data from more than 400 students (244 female, 187 male, aged 10–18) participating in online CS courses, we examined age- and gender-related patterns in enthusiasm. The findings challenge the prevailing belief that early exposure is the primary pathway to sustained interest in CS. Instead, we identify a marked decline in enthusiasm during early adolescence, particularly among girls, alongside substantial variability in interest trajectories across age groups.
Crucially, our analyses (exploratory factor methods and ANOVA) reveal that age is a more decisive factor than gender in shaping interest development and uncover key developmental breakpoints. Despite starting with lower baseline attitudes, older students showed the largest positive changes following the intervention, suggesting that well-designed short activities can effectively re-activate interest even at later ages.
Overall, the study highlights the need for a dynamic, age-sensitive framework for CS education in which instructional strategies are aligned with developmental trajectories. In addition to providing a validated instrument for capturing short-term affective and motivational responses, our research underscores the urgency of designing adaptable CS learning experiences that respond to the evolving needs of students, thereby fostering wider, more sustained engagement in CS.
\end{abstract}
\begin{document}



\flushbottom
\maketitle
%
%
\thispagestyle{empty}

\section{Introduction}
In contemporary education, computer science (CS) is recognized as a foundational discipline, essential for navigating the complexities of the digital era. However, fostering sustained engagement among a diverse student body remains a persistent challenge \cite{gorbacheva2019directions, master2025causes, royalsociety2025}. While outreach programs and workshops are frequently employed to stimulate early interest, empirical evidence suggests that student attitudes toward CS are non-linear; interest and self-efficacy often decline during the transition from childhood to mid-adolescence.

Current research increasingly advocates early introduction of computing concepts, on the premise that early exposure fosters longitudinal skill development and interest \cite{french2018females, elhamamsy2023primary, mcclure2017stem}. Nevertheless, early intervention alone may be insufficient to address the heterogeneous learning trajectories and developmental stages of a diverse cohort. This complexity is further exacerbated by systemic demographic disparities, most notably the persistent underrepresentation of female students.

This study investigates the intersection of age and gender in CS engagement to elucidate the dynamics that influence student interest. By analyzing how educational interventions modulate attitudes across different developmental stages, this research seeks to inform more targeted pedagogical strategies. Consequently, this study is guided by the following research questions: 

\begin{enumerate}
\item \textbf{Assessment Instrument Development:}
Initially, we were looking for a suitable instrument to measure the enthusiasm potential of CS interventions, one that could also detect changes in CS interest throughout an educational intervention aimed to raise engagement with the subject. As also highlighted by other researchers \cite{grosch2020mint, berg2023elaborating}, we observed a notable gap in available tools here, leading to our first research question:
\emph{How can we measure the enthusiasm potential of CS interventions?}
To address this gap, we introduce a novel questionnaire to assess the potential of CS interventions to increase interest in CS, referred to as enthusiasm potential. In Section \ref{sec:existingInstruments}, we detail our review of existing measurements and their shortcomings in the context of our research approach. During the questionnaire development process, we considered a variety of cognitive dimensions, such as positive attitudes toward computing, perceived relevance, and future intentions, based on the person-object-interest (POI) theory of interest \cite{krapp1999interest}, as well as perceptions of stereotypes and self-efficacy. The development of this instrument is detailed in Section \ref{sec:questionnaire}.
\item \textbf{Analysis of Interest Dynamics Across Age and Gender:} \emph{How do age and gender influence interest in CS?} 
Utilizing a dataset from over 400 students (ages 10–18), we employ Exploratory Factor Analysis (EFA) to identify latent interest factors and assess their stability across developmental stages. Subsequently, Analysis of Variance (ANOVA) is utilized to determine the main and interaction effects of age and gender on interest variance. This analysis aims to pinpoint critical developmental windows in which interventions most effectively enhance engagement, thereby providing a data-driven foundation for targeted pedagogical design.
\end{enumerate}

By addressing these research questions, this study aims to provide a nuanced understanding of the multifaceted factors shaping student engagement in CS and to advocate for a shift toward dynamic and diverse pedagogical frameworks. The persistent shortcomings of broad CS interventions over the past decades \cite{gorbacheva2019directions, vainionpaeae2019girls, royalsociety2025} underscore the urgent need for a methodological reevaluation. Given significant variation in how demographic groups respond to these interventions, we propose a shift toward more personalized, adaptable measurement and instructional methods. This includes addressing challenges posed by demographic specifics, such as gender differences related to puberty that appear to particularly affect girls' behavior, with long-term implications \cite{cavanagh2007puberty, marceau2011individual}.

By employing the person-object-theory of interest (POI) \cite{krapp1999interest} as a fresh analytical lens, we aim to dissect and understand variations in students' interest in CS and the potential of dedicated CS interventions to raise enthusiasm. This approach not only challenges the entrenched 'early introduction' narrative but also highlights critical yet often overlooked phases, such as early adolescence -- a period when interest in CS notably declines, especially among female students.

\section{Background}
This section provides an overview of existing research about the factors age and gender in shaping computing attitudes, as well as about theoretical foundations related to perspectives on interest, including self-determination theory and person-object theory of interest (which will be detailed further in Section \ref{sec:theoretical-bg} in the context of the theoretical framework of the questionnaire). Finally, this section addresses the limitations of existing measurement instruments for assessing enthusiasm in CS, thereby setting the stage for the introduction of a new questionnaire to address this gap.

\subsection{Age and Gender in Computer Science Education}
Exploring factors influencing attitudes toward CS has become increasingly important as the field evolves and expands across educational landscapes. Age, gender, and computer experience are recognized as critical determinants of these attitudes, though research on the effect of age on interest trajectories in CS remains sparse.

Comber and colleagues \cite{comber1997effects, colley2003age} underscored the significant influence of age and gender on computing attitudes, with younger students and males generally exhibiting more positive dispositions towards technology. They also highlighted the potential of direct computer experience to counteract negative perceptions, particularly among older and female students, emphasizing the importance of exposure and familiarity in fostering positive attitudes towards CS.

Building on these insights, recent studies have shed light on the nuanced ways gender dynamics and developmental stages intersect with CS interest. Master et al. \cite{master2021gender} found that gender-interest stereotypes more strongly affect girls' motivation to engage with CS and engineering than do gender-ability stereotypes, leading to lower interest and a diminished sense of belonging in these fields among girls compared to boys. These findings underline the pervasive role of societal stereotypes in shaping gender disparities in CS engagement.

Moreover, considerations around the timing of puberty have been linked to academic engagement and interest. Cavanagh et al. \cite{cavanagh2007puberty} highlight how the psychological risk of early puberty can translate into long-term disadvantages for girls. Similarly, Marceau et al. \cite{marceau2011individual} indicate that the timing and tempo of puberty play significant roles in the development of behavioral problems, particularly among girls, which affect their engagement in academic and extracurricular activities.

\subsection{Theories of Interest}
Interest is undoubtedly one of the most fundamental values in education \cite{harackiewicz2016interest}. In a theory-based manner, interest is widely operationalized by three approaches. Following a first theoretical approach, interest is characterized as both a desirable outcome and a multidimensional construct, defined by the relational quality between an individual and a specific object \cite{krapp2002structural}. Under the Person-Object Theory of Interest (POI) framework, an ``object'' may encompass specific activities (e.g., scientific inquiry) or discrete domains of knowledge (e.g., professional expertise). The depth of this interest is empirically measured through the individual's subjectively perceived value (value-related valences) and their qualitative emotional experience (feeling-related valences) during engagement with the object.

A second theoretical framework, Self-Determination Theory (SDT) \cite{DeciR85}, posits that positive emotional experiences are intrinsically linked to autonomous behavioral regulation. According to SDT, the development of motivation and interest is predicated on the satisfaction of three fundamental psychological needs: competence, autonomy, and social relatedness \cite{DeciR85}.

Complementing these views, a third approach emphasizes that interest is shaped by the interaction between stable personal preferences and situational environmental stimuli \cite{hidi2006four,renninger2011revisiting}. Krapp \cite{krapp2002structural} distinguishes between two manifestations of interest based on their etiology: on the one hand, if the origin of interest lies in the attractiveness of the learning object or the learning environment, it is \emph{situational interest}; on the other hand, \emph{actualized (individual) interest} means momentary interest that is caused by personal dispositions, i.e. the interest that already exists in a person. Some authors use the umbrella term ``current interest'' to refer to these two forms of short-term interest.

From a psychometric perspective, Knogler et al. \cite{knogler2015situational} further distinguish situational interest into ``catch'' and ``hold" components. The ``catch'' component refers to the initial stimulation of attention and positive affect by the environment. In contrast, the ``hold'' component involves the perceived personal significance of the content and an epistemic orientation—the desire to explore the topic further and seek new information (value-related valence).

The ``hold'' component is critical for the longitudinal maintenance of interest. While situational interest often wanes rapidly after the conclusion of a learning activity \cite{krapp2002structural}, the transition from ``catch'' to ``hold'' is a critical phase in preventing the typical decline in engagement in educational settings.

\subsection{Existing Measurement Instruments}
\label{sec:existingInstruments}
The evaluation of educational interventions is supported by several foundational works offering standardized questionnaires. These instruments primarily assess constructs such as motivation, interest, perception, and self-efficacy, with each framework typically focusing on a singular construct. Notably absent, however, are tools specifically designed to measure the enthusiasm potential of CS interventions, i.e., their potential to change students' attitudes towards the subject, underscoring the need for the development of a new questionnaire \cite{grosch2020mint, berg2023elaborating}.

The \emph{Motivated Strategies for Learning Questionnaire (MSLQ)} by Pintrich and de Groot \cite{pintrich1990motivational} is designed to assess learning strategies and academic motivation among college students. Similarly, the \emph{Science Motivation Questionnaire II (SMQ)} by Glynn \cite{glynn2011science} evaluates learning motivation within the sciences. The \emph{Individual Interest Questionnaire (IIQ)} by Rotgans \cite{rotgans2015validation}, developed to measure individual interest across various subjects and age groups, uses broadly applicable questions.

Specific to CS, the \emph{Computational Thinking Scales (CTS)} by Korkmaz et al. \cite{korkmaz2017validity}, and the \emph{Computer Programming Self-Efficacy Scale (CPSES)} by Kukul et al. \cite{kukul2017computer} assess competencies of students' computational thinking skills and programming skills, respectively. However, these instruments, designed for one-time administration, do not adequately capture general interest in CS.

Existing questionnaires for evaluating attitudes towards STEM subjects, as proposed in \cite{unfried2015development, faber2013students, erkut2005schools}, focus on mathematics, the natural sciences, and technology, but not explicitly on CS. Moreover, these instruments are generally intended for post-test administration only, limiting their ability to capture the dynamic nature of interest development throughout specific interventions.

Although these scales are often well validated with large participant pools, they fall short in measuring the potential of specific interventions to raise enthusiasm for CS -- a complex matter not fully addressed by any existing instrument. Their design typically emphasizes specific components or is confined to particular contexts. As noted by other researchers in CS education \cite{grosch2020mint, berg2023elaborating}, the absence of a standardized methodology for reliably measuring STEM interests is a significant gap. Our work seeks to address this by introducing a comprehensive approach to measuring the enthusiasm potential of CS interventions, facilitating the 

\section{Methodology}
\begin{figure}
    \centering
    \includegraphics[]{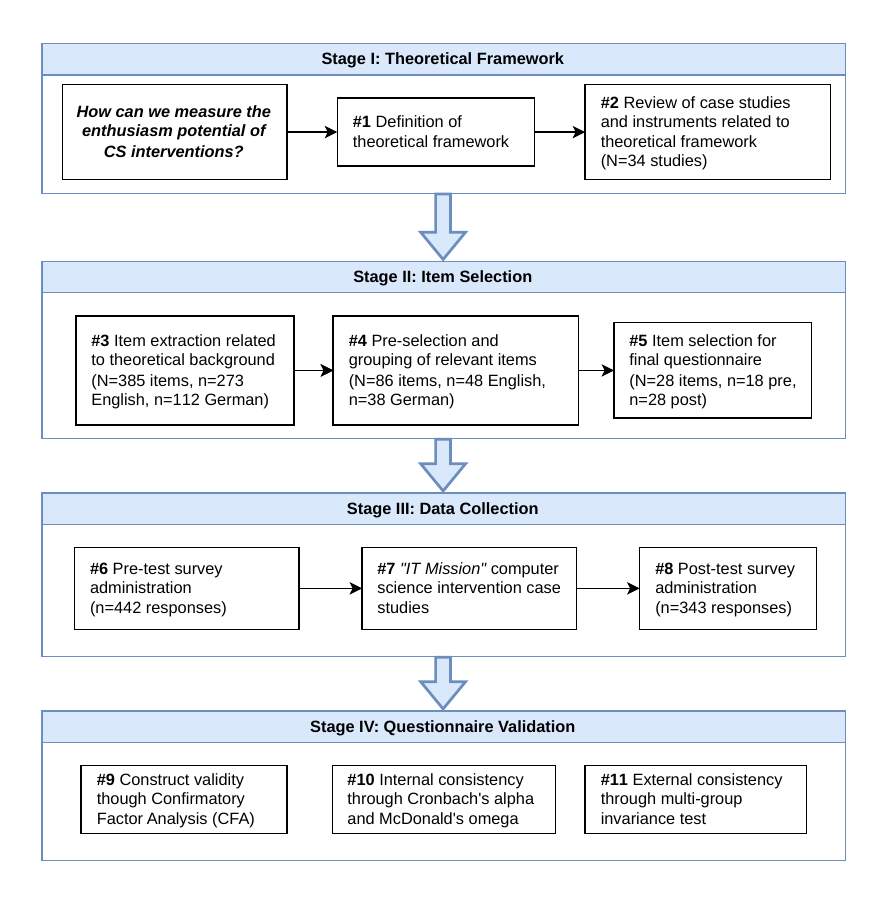}
    \caption{Schematical overview of research procedure.}
    \label{fig:process}
\end{figure}

This section delineates the research methodology, structured into four primary stages as illustrated in Figure \ref{fig:process}.

\paragraph{\textbf{Stage I \& II: Questionnaire Development}}
The initial stage involved establishing a theoretical foundation (Section \ref{sec:theoretical-bg}) and conducting a comprehensive review of existing case studies (Section \ref{sec:lit-review}). Following this, the second stage focused on extracting and selecting relevant items from existing instruments, culminating in the formulation of the final questionnaire, comprising 18 items in total (17 5-point Likert scale items and 1 open-ended item; see Section \ref{sec:FinalQuestionnaire}). Section \ref{sec:questionnaire} details the development process.

\paragraph{\textbf{Stage III: Data Collection}}
During the third stage, data were collected through several case studies involving students participating in workshops called \emph{IT Mission}, where they enrolled in online courses from the \rstname~initiative\cite{marquardt2024crossing, happe2025authentic} (freely available online: \rstwww). 

Rather than being designed for a specific age group, the main objective of the course design was to require no prior knowledge and to focus on engaging design by incorporating interdisciplinary, real-world contexts relevant to students' lives, such as addressing bee mortality by employing techniques from web development or data science, or applications of AI in the domain of music. Although not the primary objective, the courses were designed to align with the curriculum for 7th- to 10th-grade students in the study region, which covers ages 12 to 16.

The online courses were the main activity for the students during each case study. This allowed us to have very similar conditions for all case studies. Each case study followed a uniform procedure: 1) a short introduction, 2) voluntary completion of the pre-test questionnaire, 3) participation in at least one of the online courses with instructor support available, and 4) completion of the post-test questionnaire. Each study typically lasted about two hours. These studies spanned diverse educational settings, including out-of-school workshops and regular school sessions, to ensure a broad sample across educational contexts. In total, we collected more than 430 responses from secondary school students (see Table \ref{tab:frequ_age_gender}). Participants were informed about their rights to participate in the questionnaire, and participation was voluntary.

\begin{table}[]
    \centering
    \caption{Frequency of responses by age groups (AG) and gender in pre-test and post-test. Some cases did not specify gender.}
    \label{tab:frequ_age_gender}

\centering
\begin{tabular}{lrrrlrrr} 
\toprule
  Age group (AG)                & \multicolumn{3}{l}{pre-test} &  & \multicolumn{3}{l}{post-test}  \\ 
\cmidrule{2-4}\cmidrule{6-8}
 & female & male & total        &  & female & male & total          \\ 
\midrule
AG1 (10 to 12)    & 61     & 54   & 115          &  & 47     & 34   & 81             \\
AG2 (13 to 15)    & 124    & 110  & 239          &  & 104    & 71   & 177            \\
AG3 (16 to 18)    & 59     & 23   & 85           &  & 49     & 20   & 70             \\ 
\midrule
total             & 246    & 187  & 439          &  & 200    & 125  & 328            \\
\bottomrule
\end{tabular}
\end{table}

\paragraph{\textbf{Stage IV: Analysis}}
Preliminary to our main analysis, we validated our questionnaire following the procedures described by \cite{rotgans2015validation}. This involved assessing construct validity through Confirmatory Factor Analysis (CFA) \cite{scholters2011what, bollen2002latent}, evaluating internal consistency using Cronbach's alpha ($\alpha$) and McDonald's omega ($\omega$) \cite{dunn2014alpha}, and verifying structural invariance across diverse educational contexts.

Subsequently, we conducted an Exploratory Factor Analysis (EFA) using Maximum Likelihood (ML) extraction and Oblimin rotation based on parallel analysis \cite{fabrigar1999evaluating}. 
Then, two-way ANOVA was used to examine the intersectional effects of age and gender, and one-way ANOVA to examine age- and gender-dependent interest development. Despite deviations from normality assumptions, ANOVA's robustness to such violations justified its application \cite{glass1972consequences, harwell1992summarizing}. To analyze the effects of single independent variables (age or gender), we used Welch's test to validate the one-way ANOVA results \cite{kim2014analysis}. The choice of Likert-type scales in our questionnaire does not impose any restrictions on the statistical tests we selected \cite{norman2010likert}.

Because our study required a large number of statistical tests on the same sample, we applied the Bonferroni correction to adjust our significance levels, thereby controlling for the increased risk of Type I errors associated with multiple comparisons \cite{holm1979simple}. Specifically, the pre-test data comprised 17 tests (one for each item), and the post-test data included an additional 10 items, for a total of 27 tests, with a subsequent 17 tests assessing changes, totaling 34 tests for the pre-test data sample and 51 tests for the post-test data sample analyses.

To apply the Bonferroni correction, we divided the conventional alpha level of p=.05 by the number of tests conducted at each stage: 0.05/34=0.015 for pre-test analyses and 0.05/51=0.001 for post-test analyses. Accordingly, we set a corrected significance level of p$<$.001, ensuring that our results are robust against the inflation of false positive rates that could otherwise emerge from the extensive testing on the same sample set.

We also report descriptive statistics including means (M) and 95\%-confidence intervals (95\%-CI) as well as eta squared ($\eta^{2}$) for estimation of effect sizes \cite{richardson2011eta}, and statistical power indicating results' reliability \cite{cohen1988statistical}.

Statistical analyses were conducted using a suite of specialized software: \emph{SPSS} \cite{spss2025} for ANOVA, \emph{jamovi} \cite{jamovi2025} for Exploratory Factor Analysis (EFA), and the \emph{lavaan} package in R \cite{r2023} for structural equation modeling required for questionnaire validation.

\paragraph{\textbf{Ethical Compliance and Consent to Participate}}
All procedures conducted in this study adhered to the relevant institutional and national guidelines and regulations governing research involving human participants. 
Participation in the case studies and the associated questionnaires was entirely voluntary. Before completing the questionnaires, all participants were informed about the purpose of the study, the anonymous nature of the data, their right to withdraw at any time without consequence, and the intended use of results for research purposes. Informed consent was obtained from all participants. For minors, consent was obtained from their legal guardian(s) in addition to the minor's consent, in accordance with institutional guidelines. No identifying data were collected or stored, and no incentives were provided for participation.

\section{Development of the Questionnaire}
\label{sec:questionnaire}
The creation of the questionnaire mainly happened in two stages following seven steps (see Figure \ref{fig:process}). The process was guided by the question ``How to measure enthusiasm potential of CS interventions''.

\subsection{Goal Definition}
\label{sec:goaldefinition}
The primary goal of our instrument is to assess the enthusiasm potential of interventions to increase students' interest in CS. Our approach focuses on their initial interest and the impact of educational interventions on the development of attitudes and perceptions. 
Therefore, our instrument is dedicated to pre-test-post-test study designs that directly compare students' attitudes towards CS before and after participation in a specific intervention, such as workshops, lessons in school, or online interventions, such as online courses. 
Targeting students in grades 6 to 10 (ages 11 to 16), the questionnaire is designed to be comprehensible and engaging for this age group, ensuring it effectively captures their evolving perceptions and enthusiasm for CS.

\subsection{Theoretical Framework}
\label{sec:theoretical-bg}
In this first step, we defined the theoretical framework of our instrument by conceptualizing \emph{enthusiasm} in terms of cognitive constructs related to motivation and interest. We also consider subject-related perceptions as relevant variables for measuring enthusiasm potential.

\subsubsection{Enthusiasm, Motivation, and Interest}
\label{enthusiasm}
To investigate how to raise and measure enthusiasm for CS, we must first agree on a common understanding of its meaning. Despite its importance in education, there is no universally accepted definition of enthusiasm. Our first definition approach is related to Singh et al. \cite{singh2020towards} and Alpay et al. \cite{alpay2008student}, defining enthusiasm as the willingness to engage with a topic. As such, more established cognitive constructs, such as motivation and interest, are closely related to enthusiasm \cite{wommel2016enthusiasmus}. Motivation can be classified as extrinsic or intrinsic \cite{gopalan2017review}. On the one hand, intrinsic motivation is a strong predictor of enthusiasm, as studies have observed strong correlations between intrinsic motivation and career orientation \cite{aivaloglou2019early, glynn2011science}, consistent with the view of enthusiasm as the willingness to engage with a topic. On the other hand, there is the construct of interest, which is barely separable from the construct of intrinsic motivation \cite{krapp1999interest}.

The conceptualization of the construct interest is very much shaped by the researchers Krapp, Prenzel, and Schiefele, concluding in the Person-Object-Theory of Interest (POI) \cite{krapp2007educational, schiefele1991interest, schiefele1983principles}. The theory describes interest as a positive relationship between a person and an object. In that context, an object does not necessarily have to be a concrete physical thing, instead in educational contexts it usually rather refers to e.g., a topic, subject matter, or an abstract idea \cite{krapp2007educational}. According to the POI, there are essentially three underlying characteristics of interest \cite{pawek2009schulerlabore}: positive feelings (i.e., emotional characteristics \cite{krapp2002structural, krapp2007educational}), relevance (i.e,. value-related characteristics \cite{krapp2007educational}), and future intents (i.e,. epistemic orientation \cite{pawek2009schulerlabore, knogler2015situational}):
\begin{itemize}
    \item \textsc{\textbf{Positive Feelings}}: This characteristic of interest relates to the extent to which an activity is connected to positive emotions, such as fun \cite{krapp2002structural}. In the best case, this results in a \emph{flow} momentum \cite{csikszentmihalyi1990flow} in which ``time flies by''.
    \item \textsc{\textbf{Relevance}}: This characteristic represents the self-intentionality and self-identification of a relationship between a person and an object expressed by the extent to which ``goals and volitionally realized intentions related to the area of an interest are compatible with the attitudes, expectations, values and other aspects of the person’s self-system'' \cite[p. 11]{krapp2007educational}. Essentially, the object of interest is considered personally relevant, and the person can identify himself with the object.
    \item \textsc{\textbf{Future Intents}}: The third characteristic is reflected in a person's desire, or development of such a desire, to expand their competencies concerning the subject of interest, to increase their knowledge, and to improve their skills \cite{pawek2009schulerlabore, knogler2015situational}.
\end{itemize}

The POI aligns well with the concept of \emph{subject enthusiasm} highlighted by Kunter et al. in their study on teacher enthusiasm \cite{kunter2011teacher}. In their work, the authors differentiate between \emph{teaching enthusiasm} and \emph{subject enthusiasm}. While the first case is not relevant to our approach, the second is, as we aim to measure the potential of interventions to increase participating students' enthusiasm for a specific subject (i.e., computer science). According to the authors, the concept of subject enthusiasm can be expressed by the POI to a high degree, particularly its emotional component \cite{kunter2011teacher}.

Considering our remarks and combining the different enthusiasm approaches, we understand the enthusiasm potential of an intervention as its potential impact on students' \emph{thirst for action and willingness to engage with a learning object} expressed by changes in the three dimensions of interest (positive feelings, relevance, future intents) derived from the theoretical model of the Person-Object-Theory of Interest (POI).

\subsubsection{From Situational Interest to Individual Interest}
Consideration in the investigation of interest necessitates the distinction between situational and individual interest \cite{krapp1999interest}. \emph{Situational interest} can be understood as a current and time-limited state that arises from a specific situation and is often bound to that situation. \emph{Individual interest}, on the other hand, describes a long-term developmental outcome. Therefore, it is also referred to as \emph{long-term interest} or \emph{sustained interest} and is often used synonymously with the basic term ``interest''.

A specific intervention typically elicits current emotions about a subject of interest within the context of the intervention for a limited time. 
This can be demonstrated, for example, by a person rating a topic much more positively when they have just been intensively engaged with it, compared to a later point in time, when it is eventually overshadowed by other influences. In this context, two aspects of interest warrant investigation: first, the effect of an intervention on situational and updated interest; and second, its impact on long-term individual interest. While the effect on long-term interest requires extensive long-term studies, dynamic changes in situational interest can be relatively well captured within limited pre-test-post-test study designs tailored to specific, time- and space-limited interventions (such as a single workshop or a single school lesson).

However, it is important to note that those two concepts of interest are mutually interconnected as situational interest plays a central role in the development of individual interest \cite{rotgans2017interest}. Especially when it occurs regularly, significant and positive influences on a person's general and long-term individual interest have been observed for situational interest \cite{palmer2017using, friend2017girls}.

\subsubsection{Subject-related Perceptions}
Perceptions of a subject and subject-related activities influence attitudes towards it and thus play a relevant role in developing subject-specific interest. It is well known that, in CS, several distinct (mis-)conceptions and stereotypes exist that can negatively affect the perception of the subject among certain groups of learners and create barriers to interest development \cite{master2021gender, beyer2014women}. This is why observing how perception changes during an intervention can be highly beneficial for gaining further insights into a learner's overall development of subject-specific interests.

In terms of CS, stereotypes related to gender (e.g., CS is a field for men), personalities (e.g., CS is done by nerds), and environment (e.g., CS is done in a dark basement), among others, are known to negatively impact the attitude towards CS of certain groups of students \cite{master2021gender}. But other perceptions also play relevant roles in potential interest development, such as the belief that CS is only about programming, with little or no practical relevance and disconnected from students' everyday life experiences \cite{cheryan2015cultural, lewis2016don}. 
Self-efficacy is another relevant perception-related construct used to assess an individual's self-confidence. Especially in STEM subjects, it is well known that strengthening self-efficacy and challenging stereotypes are pivotal for sparking enthusiasm and fostering interest in CS \cite{beyer2014women, happe2021effective, aivaloglou2019early, tellhed2022sure}.

\subsection{Review of Case Studies}
\label{sec:lit-review}
As a starting point for our review of case studies, we used a comprehensive literature review by Happe et al. \cite{happe2021effective} about measures and strategies for fostering interest in CS among girls. This provided us with a solid body of relevant case studies in the context of increasing and fostering interest in CS.
In addition, we expanded our search using Google Scholar. Several queries have been conducted, with each query being a combination of one or more of the following keywords (in English and German):
\begin{itemize}
    \item \emph{demographic:} gender, girls, women
    \item \emph{subject:} stem, computer science
    \item \emph{educational:} course, intervention, school, teach*, edu*
    \item \emph{type:} e-learning, mooc, online, digital, interdisciplinary
    \item \emph{concept:} interest, motivation, attitude, enthusiasm, engagement, excitement
    \item \emph{misc:} evaluation, framework, measure, scale, questionnaire
\end{itemize}

In the selection process, we applied specific inclusion criteria aligned with our research objective, focusing on studies that investigate predefined constructs relevant to our theoretical framework, employ pre-test and post-test surveys, provide access to the surveys, focus on CS, and target students aged 10 to 18. Conversely, we excluded studies that were not aligned with our theoretical framework, lacked publicly accessible survey instruments, or were not available in English or German.
Table \ref{tab:table-lit} shows the final dataset, which included 33 studies (28 English, 5 German).

\begin{table}[]
    \centering
    \caption{Overview of studies from the literature review providing instruments relevant to the theoretical framework of this study.}
    \label{tab:table-lit}

\begin{tabular}{lcclllccc} 
\toprule
study                                           & \begin{sideways}language\end{sideways} & \begin{sideways}interest\end{sideways} & \multicolumn{1}{c}{\begin{sideways}perception\end{sideways}} & \multicolumn{1}{c}{\begin{sideways}self-efficacy\end{sideways}} & \multicolumn{1}{c}{\begin{sideways}sterotypes\end{sideways}} & \begin{sideways}pre \& post\end{sideways} & \begin{sideways}\begin{tabular}[c]{@{}c@{}}\#items\\\end{tabular}\end{sideways} & \begin{sideways}age range\end{sideways}  \\ 
\midrule
Beumann \cite{beumann2017versuch}                      & ger                                    & x                                      &                                                              &                                                                 &                                                              & yes                                     & 32                                                                              & n.a.                                     \\
Blankenburg et al. \cite{blankenburg2015naturwissenschaftliche}       & ger                                    & x                                      & \multicolumn{1}{c}{x}                                        &                                                                 &                                                              & no                                      & 76                                                                              & 10-18                                    \\
Burns \& Lesseig \cite{burns2017}               & en                                     & x                                      &                                                              &                                                                 &                                                              & yes                                     & n.a.                                                                            & 10-15                                    \\
Chipman et al. \cite{chipman2018evaluating}               & en                                     & \multicolumn{1}{l}{}                   &                                                              & \multicolumn{1}{c}{x}                                           &                                                              & yes                                     & 15                                                                              & 10-18                                    \\
DuBow \& J.-Hawkins \cite{dubow2016male}            & en                                     & x                                      &                                                              & \multicolumn{1}{c}{x}                                           &                                                              & no                                      & 38                                                                              & $>$18                                       \\
Ericson \& McKlin \cite{ericson2012effective}            & en                                     & x                                      & \multicolumn{1}{c}{x}                                        & \multicolumn{1}{c}{x}                                           & \multicolumn{1}{c}{x}                                        & yes                                     &  $>$8                                                                              & 6-18                                     \\
Erkut \& Marx \cite{erkut2005schools}                  & en                                     & x                                      & \multicolumn{1}{c}{x}                                        & \multicolumn{1}{c}{x}                                           &                                                              & yes                                     & \textasciitilde{} 40                                                            & n.a.                                     \\
Ertl et al. \cite{ertl2014stereotye}                     & ger                                    & x                                      &                                                              &                                                                 & \multicolumn{1}{c}{x}                                        & no                                      & 28                                                                              & n.a.                                     \\
Faber et al. \cite{faber2013students}                   & en                                     & x                                      & \multicolumn{1}{c}{x}                                        &                                                                 &                                                              & no                                      & \textasciitilde{} 50                                                            & 6-18                                     \\
Friend \cite{friend2017girls}                        & en                                     & x                                      & \multicolumn{1}{c}{x}                                        &                                                                 &                                                              & yes                                     &  $>$5                                                                              & 10-18                                    \\
Glowinski \cite{glowinski2007schuelerlabore}                  & ger                                    & x                                      &                                                              &                                                                 &                                                              & no                                      & 55                                                                              & 16-18                                    \\
Glynn et al. \cite{glynn2011science}                   & en                                     & x                                      &                                                              & \multicolumn{1}{c}{x}                                           &                                                              & no                                      & 25                                                                              & $>$18                                       \\
Haselmeier et al. \cite{haselmeier2019interesse}         & ger                                    & x                                      &                                                              & \multicolumn{1}{c}{x}                                           &                                                              & no                                      & n.a.                                                                            & 10-15                                    \\
Häußler \cite{haussler2007lasst}                    & ger                                    & x                                      &                                                              &                                                                 &                                                              & yes                                     & \textasciitilde{} 20                                                            & n.a.                                     \\
Henry \& Dumas \cite{henry2018}                 & en                                     & x                                      &                                                              &                                                                 &                                                              & no                                      & 4                                                                               & 10-15                                    \\
Holmes et al. \cite{holmes2018integrated}                 & en                                     & x                                      &                                                              &                                                                 &                                                              & no                                      & n.a.                                                                            & $>$10                                       \\
Jenson \& Black \cite{jenson2017gender}               & en                                     & x                                      & \multicolumn{1}{c}{x}                                        &                                                                 & \multicolumn{1}{c}{x}                                        & yes                                     & \textasciitilde{}20                                                             & 10-15                                    \\
K.-Hallak et al. \cite{kaloti2015students}              & en                                     & x                                      & \multicolumn{1}{c}{x}                                        & \multicolumn{1}{c}{x}                                           &                                                              & yes                                     & 31                                                                              & 10-15                                    \\
Katterfeldt et al. \cite{katterfeldt2019effects}       & en                                     & x                                      &                                                              &                                                                 &                                                              & yes                                     & 22                                                                              & 10-18                                    \\
Kirikkaya \cite{kirikkaya2011grade}                  & en                                     & x                                      & \multicolumn{1}{c}{x}                                        &                                                                 & \multicolumn{1}{c}{x}                                        & no                                      & 43                                                                              & 10-18                                    \\
Kukul et al. \cite{kukul2017computer}                   & en                                     & \multicolumn{1}{l}{}                   &                                                              & \multicolumn{1}{c}{x}                                           &                                                              & no                                      & 31                                                                              & 10-15                                    \\
Master et al. \cite{master2017}                 & en                                     & x                                      &                                                              &                                                                 & \multicolumn{1}{c}{x}                                        & yes                                     &  $>$4                                                                              & 6-10                                     \\
Müller et al. \cite{muller2007skalen}                 & ger                                    & x                                      &                                                              &                                                                 &                                                              & no                                      & 17                                                                              & 10-18                                    \\
Ng \& Fergusson \cite{ng2020engaging}                   & en                                     & x                                      &                                                              & \multicolumn{1}{c}{x}                                           &                                                              & yes                                     & 30                                                                              & 10-15                                    \\
Outlay et al. \cite{outlay2017getting}                 & en                                     & x                                      & \multicolumn{1}{c}{x}                                        &                                                                 & \multicolumn{1}{c}{x}                                        & yes                                     &  $>$7                                                                              & 10-15                                    \\
Palmer et al. \cite{palmer2017using}                 & en                                     & x                                      &                                                              &                                                                 &                                                              & yes                                     & \textasciitilde{} 14                                                            & $>$ 18                                       \\
Pintrich \& de Groot \cite{pintrich1990motivational}        & en                                     & \multicolumn{1}{l}{}                   &                                                              & \multicolumn{1}{c}{x}                                           &                                                              & no                                      & 44                                                                              & 10-15                                    \\
Rotgans \cite{rotgans2015validation}                      & en                                     & x                                      &                                                              &                                                                 &                                                              & yes                                     & 19                                                                              & 10-15                                    \\
Sabin et al. \cite{sabin2017}                   & en                                     & x                                      &                                                              & \multicolumn{1}{c}{x}                                           &                                                              & yes                                     & 22                                                                              & 10-18                                    \\
Schorr \cite{schorr2019pipped}                        & en                                     & x                                      &                                                              & \multicolumn{1}{c}{x}                                           &                                                              & no                                      & \textasciitilde{} 50                                                            & 10-18                                    \\
Theodoropoulos et al. \cite{theodoropoulos2018computing} & en                                     & x                                      &                                                              & \multicolumn{1}{c}{x}                                           &                                                              & yes                                     & 24                                                                              & 10-18                                    \\
Unfried et al. \cite{unfried2015development}               & en                                     & x                                      & \multicolumn{1}{c}{x}                                        & \multicolumn{1}{c}{x}                                           &                                                              & no                                      & 94                                                                              & 6-18                                     \\
Vela et al. \cite{vela2018matters}                     & en                                     & x                                      & \multicolumn{1}{c}{x}                                        &                                                                 &                                                              & yes                                     &  $>$11                                                                             & 16-18                                    \\
\bottomrule
\end{tabular}
\end{table}

\subsection{Item Selection}
\label{sec:item-selection}
In the first selection step, the questionnaires used in each case study were examined separately. For each study, all items related to one of the constructs defined in \ref{sec:theoretical-bg} were extracted into a single comprehensive Excel file, retaining information on the item source, goal variable, scale type, and age range. That is, if a study using the questionnaire also included questions about, e.g., skills or competencies, those questions were excluded from the extraction process, whereas other relevant items from the same questionnaire could have been included. This first step produced a dataset of 385 items (273 English, 112 German).

In the next step, items from the dataset were systematically selected and grouped by subject of investigation and question type. Initially, items addressing the same subject of investigation (e.g., stereotypes) and having similar formulations were assigned to the same group. For each group of items, one item was selected as the representative. 

In the following, the approach is exemplified on the item \emph{``I enjoy solving problems with computers''} (see Table \ref{tab:item-group}).

\begin{table*}
\centering
  \caption{The group of questions on the item ``\emph{I enjoy solving problems with computers}.´´}
\label{tab:item-group}
\begin{tabular}{lll}
\# & Item text & Source \\ \toprule
R & \textbf{I enjoy solving problems with computers}        &                         \\ \midrule
1 & I enjoy solving problems with computers & \cite{chipman2018evaluating} \\
2 & Computers are fun         &    \cite{outlay2017getting}                                                   \\
3 & Computers are fun to use   &             \cite{chipman2018evaluating}                                        \\
4 & How fun is programming?   &   \cite{master2017}                                                   \\
5 & I like computing     &        \cite{ericson2012effective}                                                   \\ 
6 & How fun are robots?     &      \cite{master2017}                                                  \\
7 & Ich arbeite und lerne in diesem Fach, & \cite{muller2007skalen} \\ 
 & weil ich gerne Aufgaben aus dem Fach löse & \\ \bottomrule

\end{tabular}
\end{table*}

Although the questions may seem very different at first glance, they were grouped according to clear criteria aligned with the framework's objectives. All questions aim to determine whether and to what extent someone enjoys performing a typical subject-related task. The representative was chosen because the question directly references CS and is neither too specific (like item 4) nor too general (like item 2). Selecting a representative does not preclude using multiple questions from the same group in an alternative questionnaire.

If the representative item was in English, it was translated into German. In an iterative process, several alternative formulations were developed for each representative to ensure the item's comprehensibility while aligning with the research objective. At the end of this process, a total of 86 items (48 English, 38 German) were considered.

In the final two steps, individual items were selected for inclusion in the final questionnaire. The selection was based on the item's relevance to the instrument's objectives. Consideration was given to whether and to what extent an item evaluates one of the constructs from our theoretical framework. Additionally, the approaches and results of the studies from which the item was extracted were considered indicators. 

A limiting factor in the selection process was the total number of questions. An assessment had to be made of the questionnaire size and the time frame. To minimize barriers to completing the questionnaire, a benchmark of five to ten minutes was set for completion time. This factor may be negligible in in-person evaluations. However, when using the instrument in an online setting, the questionnaire must be easily accessible so that students are willing to complete it voluntarily and without supervision.

The initial versions of the pre-test and post-test questionnaires contained 23 and 34 questions, respectively. Following a review of the items within the research team, five questions were removed from the pre-test questionnaire and one from the post-test questionnaire.

\subsection{Final Questionnaire}
\label{sec:FinalQuestionnaire}
Table \ref{tab:cseis} shows the items of the final questionnaire consisting of 18 items that are identically repeated in the pre-test and post-test for capturing changes in students' general attitude towards CS and an additional 10 items that are only included in the post-test to capture students' feelings directly related to activities in the intervention (``course-related attitudes''). 
The source column indicates the origin from which an item was derived. The original questionnaire is in German and is available from the authors upon request. The English versions of items shown in Table \ref{tab:cseis} were translated for publication and did not undergo a standardized translation process. If items were identical to their original English resources, the original item text was used.

Most of the items are measured using a common Likert scale ranging from (1) -- Strongly Disagree, (2) -- Tend to Disagree, (3) -- Neutral, (4) -- Tend to Agree, to (5) -- Strongly Agree. This scale was chosen to facilitate questionnaire usability, given the primary target audience of secondary school students (aged 11-16), to minimize cognitive load and prevent respondent fatigue that might lead to random responses on later items. For an older adolescent audience, a 7-point scale may be appropriate to provide more nuanced response options and more precisely reflect their perceptions \cite{joshi2015likert}.

The item \emph{TX\_PE5\_CSEverywhere} is scaled from (1) -- Only in very specific fields to (5) -- Just everywhere, while \emph{TX\_PE3\_Words} is an open-ended item, allowing respondents to freely express their associations with the subject.

\begin{landscape}
\begin{table*}
    \centering
    \caption{Items included in the final questionnaire. 
    \emph{TX} is replaced with \emph{T1} for pre-test assessment and \emph{T2} for post-test assessment.
    }
\resizebox{0.95\linewidth}{!}{
\centering
\setlength{\extrarowheight}{0pt}
\addtolength{\extrarowheight}{\aboverulesep}
\addtolength{\extrarowheight}{\belowrulesep}
\setlength{\aboverulesep}{0pt}
\setlength{\belowrulesep}{0pt}
\begin{tabular}{llll} 
\toprule
Classification                                                          & Item                                                        & Item Text                                                                                                                   & Source                                                                                   \\ 
\toprule
\multicolumn{4}{l}{\textbf{General Attitudes toward CS (pre-test and post-test)}}                                                                                                                                                                                                                                                                                           \\ 
\toprule
\multirow{3}{*}{Positive Feelings (PF)}   & \cellgrey TX\_PF1\_ExerciseComp   & \cellgrey I enjoy solving problems with computers                                                 & \cellgrey \cite{chipman2018evaluating}        \\
                                                                         & TX\_PF2\_LearningComp                                       & Learning about what computers can do is fun                                                                                 & \cite{friend2017girls}                                                  \\
                                     & \cellgrey TX\_PF3\_CodingFun   & \cellgrey Coding is fun for me                                                 & \cellgrey \cite{beumann2017versuch}        \\ 
\midrule
\multirow{5}{*}{Relevance (RE)}                                               & TX\_RE1\_CSInterest                                         & I am interested in computer science                                                                                         & \cite{beumann2017versuch}                                               \\ 
                                   & \cellgrey TX\_RE2\_CodingEveryday & \cellgrey Coding skills can help me at my everyday life                                           & \cellgrey \cite{theodoropoulos2018computing}  \\
                                                                      & TX\_RE3\_CareerRelevance                                    & What I learn in computer science I know I can put to good use later on                                                      &  \cite{muller2007skalen}                                                 \\
                                     & \cellgrey TX\_RE4\_CSJobsBoring   & \cellgrey Computing jobs are boring                                                               & \cellgrey \cite{ericson2012effective}         \\
                                                                      & TX\_RE5\_CSTopicsInteresting                                & Computer scientists deal with interesting topics                                                                            &        \cite{schorr2019pipped}                                                                                  \\
\midrule
\multirow{3}{*}{Future Intents (FI)}      & \cellgrey TX\_FI1\_NoCoding       & \cellgrey I do not want to deal with coding in my life                                            & \cellgrey \cite{theodoropoulos2018computing}  \\
                                                                        & TX\_FI2\_LearningMore                                       & I would be interested in learning more about computer science than I need for school                                        & \cite{palmer2017using}                                                  \\
                                   & \cellgrey TX\_FI3\_Career         & \cellgrey I can see myself doing something in the field of computer science later on after school & \cellgrey  \cite{friend2017girls}                                                    \\ 
\midrule
Self-efficacy (SE)                        & TX\_SE\_SelfEfficacy    & I know I can do well in computer science                                                & \cite{unfried2015development}       \\ 
\midrule
Stereotype (ST)                                                              & \cellgrey TX\_ST\_Stereotypes                                         & \cellgrey Computer science is an appropriate subject for both boys and girls                                                          & \cellgrey \cite{jenson2017gender}                                                 \\ 
\midrule
\multirow{5}{*}{Perception (PE)}                                               & TX\_PE1\_Interdisciplinary                                  & I like to combine knowledge from different fields to solve problems                                                         & \cite{ng2020engaging}                                                   \\ 
                                                                         & \cellgrey TX\_PE2\_CSJobsKnowing                                      & \cellgrey I know what computer science is and what computer scientists do                                                             & \cellgrey \cite{chipman2018evaluating}                                            \\
                                     & TX\_PE3\_Words                                              & What spontaneously comes to your mind about computer science? Name up to 5 keywords.                                        & \cite{katterfeldt2019effects}                                           \\
                                   & \cellgrey TX\_PE4\_CSOnlyCoding   & \cellgrey Computer scientists mainly deal with programming                                        & \cellgrey \cite{katterfeldt2019effects}       \\
                                    &  TX\_PE5\_CSEverywhere   & Computer science is... only in very specific fields or just everywhere?              &  \cite{katterfeldt2019effects}       \\
      
\bottomrule
\multicolumn{4}{l}{\textbf{Course-related Attitudes (post-test-only)}}                                                                                                                                                                                                                                                                                     \\ 
\toprule
\multirow{4}{*}{Positive Feelings (PF)}                                      & \cellgrey T2C\_PF1\_Fun         & \cellgrey It was fun to engage with the topics covered in the course                              & \cellgrey \cite{beumann2017versuch}           \\
                                                                        & T2C\_PF2\_Curiosity                                       &  The course has aroused my curiosity                                                                                         &  \cite{beumann2017versuch}                                               \\
                                    & \cellgrey T2C\_PF3\_Time        & \cellgrey During the course time flew by                                                          & \cellgrey \cite{zehren2009forschendes}        \\
                                    & T2C\_PF4\_SchoolFun                                         & School would be more fun if we would cover things like this more often                                                      & \cite{haussler2007lasst}                                                \\  
\midrule
\multirow{2}{*}{Relevance (RE)}            & \cellgrey T2C\_RE1\_Recom       & \cellgrey I would recommend such a course to others                                               & \cellgrey                                                        \\
 & T2C\_RE2\_LearningRelevance                                 &  I felt like I had learned something for myself                                                                              & \cite{haussler2007lasst}                                                \\
                                    
\midrule
\multirow{3}{*}{Future Intents (FI)}     & \cellgrey T2C\_FI1\_Repeat      & \cellgrey I would love to do a course like this again                                             & \cellgrey \cite{glowinski2007schuelerlabore}                                                    \\
                                                                       & T2C\_FI2\_Aha                                             & During the course I had an aha moment                                                                                       &  \cite{beumann2017versuch}                                               \\
                                    & \cellgrey T2C\_FI3\_Talk        & \cellgrey I will talk to friends, parents, or siblings about things I experienced in the course   & \cellgrey \cite{engeln2004schulerlabors}      \\ 
\midrule
Individual Interest (IIN)                                                             & T2C\_IIN\_InterestIncrease                                  & My interest in computer science has increased since I took the course                                                       & \cite{haussler2007lasst}                                                \\
\bottomrule
\end{tabular}
}
    \label{tab:cseis}
\end{table*}
\end{landscape} 

\subsection{Validation of the Questionnaire}
\label{sec:reliability}
To validate our questionnaire, we followed a process similar to that in \cite{rotgans2015validation}. This included evaluating construct validity via Confirmatory Factor Analysis (CFA), assessing internal reliability, and verifying structural invariance across different educational contexts. For the cross-contextual invariance analysis, we selected two distinct samples from our dataset (see Table \ref{tab:validation_descriptives}): a formal setting with seven school classes and an informal setting with nine out-of-school programs. Overall, the results provide empirical support for the validity and reliability of the questionnaire:

\begin{table}
\centering
\caption{Descriptive statistics of samples.}
\begin{tabular}{clccc}
\toprule
Sample & Type & \emph{N} & Gender & Age (SD) \\
\midrule
1 & Formal (school lesson)                      & 120               & 50\% female                                            & 13.0 (1.23)             \\
2 & Informal (weekly workshop)                        & 93                & 33\% female                      & 13.4 (1.10)             \\
\bottomrule
\end{tabular}
\label{tab:validation_descriptives}
\end{table}

\subsubsection*{Construct Validity and Model Fit}
Construct validity assesses the degree to which observed variables (questionnaire items) consistently measure their respective latent constructs \cite{scholters2011what, bollen2002latent} -- in our case, positive feelings, relevance, and future intentions. We performed CFA using Maximum Likelihood (ML) estimation to evaluate the absolute and incremental fit of our theoretical model.

As shown in Table \ref{tab:validiation_cfa}, the CFA results indicate that the constructs are well-represented in both samples. While the RMSEA values were slightly above the preferred 0.08 threshold, they remained within an acceptable range for educational research \cite{schermelleh2003evaluating}. The relative Chi-square ($\chi^{2}/df$) and CFI values near 0.90 confirm an acceptable model fit \cite{rigdon1996cfi,bentler1990comparative}. Furthermore, all factor loadings were statistically significant, confirming that the items are meaningful indicators of their respective latent variables.

\begin{table}[]
    \centering
    \caption{Results of confirmatory factor analysis.}
    \begin{tabular}{ll}
\toprule
 Sample    & Model Fit Indices \\
 \midrule 
 Formal    & $\chi^{2}$=71.82 (Scaled), df=41, $\chi^{2}/$df=1.75, p=.002 \\
            & RMSEA=0.082 (Scaled: 0.087), 90\%-CI: [0.052, 0.119] \\
            & CFI=0.922 (Scaled: 0.935) \\
Informal   & $\chi^{2}$=70.48 (Scaled), df=41, $\chi^{2}/$df=1.72, p=.003 \\
            & RMSEA=0.090 (Scaled: 0.095), 90\%-CI: [0.055, 0.132]\\
            & CFI=0.906 (Scaled: 0.914) \\ 
            \bottomrule
\end{tabular}
    \label{tab:validiation_cfa}
\end{table}

\subsubsection*{Reliability and Internal Consistency}
Reliability estimates the extent to which an instrument is free from measurement error, typically assessed via internal consistency \cite{kimberlin2008validity}. We utilized Cronbach’s alpha ($\alpha$) and McDonald’s omega ($\omega$) to verify that items within a latent variable were sufficiently correlated \cite{dunn2014alpha}.

The results in Table \ref{tab:quest-reliability} demonstrate strong internal consistency for the overarching interest construct, with indices exceeding 0.80. The sub-dimensions (PF, RE, and FI) also fell within the acceptable-to-good range (0.70$<\alpha$/$\omega>$0.90), confirming the reliability of the questionnaire in measuring the intended psychological dimensions without significant redundancy \cite{tavakol2011making}.

\begin{table}[]
    \centering
    \caption{Results of reliability tests.}
   \begin{tabular}{llllllll}
\toprule
 Sample     &     & Items & Cronbach's $\alpha$ & McDonald's $\omega$ \\
 \midrule 
 Formal & Pre-test   & all & 0.87 & 0.89  \\
        & & PF\{1,2,3\} &  0.80 &  0.80  \\
        & & RE\{1,2,3,4,5\} & 0.71 & 0.73 \\
        & & FI\{1,2,3\} & 0.61 &   0.68  \\
        & Post-test   & all & 0.87 & 0.89  \\
        & & C\_PF\{1,2,3\} &  0.80 &  0.80  \\
        & & C\_RE\{1,2,3,4,5\} & 0.71 & 0.73 \\
        & & C\_FI\{1,2,3\} & 0.61 &   0.68  \\
        \midrule
 Informal  & Pre-test   & all & 0.88 & 0.89 \\
        & & PF\{1,2,3\} &  0.68 &  0.69  \\
        & & RE\{1,2,3,4,5\} & 0.71 & 0.73  \\
        & & FI\{1,2,3\} & 0.73 &   0.74 \\
        & Post-test   & all & 0.88 & 0.89 \\
        & & C\_PF\{1,2,3\} &  0.68 &  0.69  \\
        & & C\_RE\{1,2,3,4,5\} & 0.71 & 0.73  \\
        & & C\_FI\{1,2,3\} & 0.73 &   0.74 \\
            \bottomrule
\end{tabular}
    \label{tab:quest-reliability}
\end{table}

\subsubsection*{Multi-group Invariance and Contextual Stability}
To establish the external validity of our questionnaire across diverse settings, we conducted a multi-group invariance test between the formal and informal samples \cite{rotgans2015validation}. This procedure tests whether the relationship between observed variables and latent factors remains consistent across groups.

The results of the Chi-square difference test ($\Delta\chi^{2}$=9.34, $\Delta$df=8, see Table \ref{tab:validation_multigroup}) indicate that the constrained model does not fit significantly worse than the unconstrained model. This confirms metric invariance, suggesting that our questionnaire is a valid instrument for comparing student interest across both formal school environments and informal outreach contexts.

\begin{table}[]
    \centering
    \caption{Results of the test for multi-group invariance between formal and informal education.}
    \begin{tabular}{llllll}
\toprule
  Model   & $\chi^{2}$ & df & $\Delta\chi^{2}$ & $\Delta$df & Statistical Significance  \\
  \midrule
  Unconstrained model   & 142.31 & 82 & - & - & \\
  Constrained model & 151.65 & 90 & 9.34 & 8 & n.s. \\
  \bottomrule
\end{tabular}
    \label{tab:validation_multigroup}
\end{table}

\section{Findings}
\label{sec:results}
In this section, we first present findings from exploratory factor analysis (EFA), revealing variation in factor loadings across age groups. Then we present further insights from ANOVA analysis of how the general attitude toward CS based on pre-test data differs among gender and age groups (AGs) followed by a closer look at how attitude changes as well as course-related feelings differ by age groups.

\subsection{Exploratory Factor Analysis}

In the presented exploratory factor analysis (EFA) we leave all items included, even those that have very poor factor loading to illustrate how the specific item loading is evolving with children's age. Therefore, we did not aim to optimize model fit by deleting items; we retained them in the model even when loadings were below the recommended 0.3. 

The constructed EFA models exhibited acceptable uniqueness, with means above the suggested 0.6 threshold, indicating that a singular factor model with 16 items is robust even in smaller samples \cite{maccallum1999sample}. This assertion is supported by satisfactory Bartlett's Test of Sphericity, KMO test scores (above 0.6, with an exception for boys AG1 at 0.30), and significant Chi-square fit for all models except AG1, despite minor deviations in RMSEA and TLI indices (see Table \ref{tab:efa}).

The overall model for all students identified three factors. Surprisingly, when we performed more detailed analyses only for boys and only for girls in three different age groups, only one factor was identified by the model, except for girls in AG3, where the model identified two factors.

The core of this factor, represented by stable item loading across age and gender, includes items related to career prospects in CS (\emph{T1\_FI3\_Career}), enjoyment of computing-related activities (\emph{T1\_PF2\_LearningComp}, \emph{T1\_FI2\_LearningMore}, \emph{T1\_PF3\_CodingFun}), and self-efficacy in CS (\emph{T1\_SE\_SelfEfficacy}). However, as children age, the loading of certain items shifts. Notably, the loading of item \emph{T1\_RE1\_CSInterest} increases from AG1 to AG2, and the loading of \emph{T1\_PF1\_ExerciseComp} seems to increase with age, whereas the factor loading for perceptions of CS as a field of interest (\emph{T1\_RE5\_CSTopicsInteresting}, \emph{T1\_RE4\_CSJobsBoring}) and its future relevance (\emph{T1\_RE3\_CareerRelevance}) diminish. While the factor loading for the perception of CS as a broad field (\emph{T1\_PE5\_CSEverywhere}) declined sharply with age for boys, it increased for girls. Furthermore, gender-specific trends in stereotypes about CS (\emph{T1\_ST\_Stereotypes}) reveal a notable shift for boys, moving from negative to positive loadings from AG1 to AG2, whereas, for girls, their loading for this factor diminishes with age.

\begin{landscape}
\begin{table*}[]
    \centering
    \caption{Results from exploratory factor analysis. Arrows indicate an increase (\apos, \appos), decrease (\adown, \addown), or inversion (\aupdown) of factor loading of specific items compared to their factor loading in the previous age group (AG3 $\rightarrow$ AG2 $\rightarrow$ AG1). Factor loadings below 0.30 are colored grey.}
    \label{tab:efa}

\resizebox{\linewidth}{!}{%
\begin{tabular}{lrrrlrrlrrlrrlrrlrr} 
\toprule
                                         & \multicolumn{3}{l}{All (2 factors)}                                                                                              &                      & \multicolumn{5}{l}{Boys}                                                                                                                                                                         &                      & \multicolumn{8}{l}{Girls}                                                                                                                                                                                                                                                                                        \\ 
\cmidrule{2-4}\cmidrule{6-10}\cmidrule{12-19}
                                         & \multicolumn{3}{l}{}                                                                                                             &                      & \multicolumn{2}{l}{AG1 (1 factor)}                                                  &                      & \multicolumn{2}{l}{AG2 (1 factor)}                                                  &                      & \multicolumn{2}{l}{AG1 (1 factor)}                                                  &                      & \multicolumn{2}{l}{AG2 (1 factor)}                                                   &                      & \multicolumn{2}{l}{AG3 (1 factor)}                                                    \\ 
\cmidrule{6-7}\cmidrule{9-10}\cmidrule{12-13}\cmidrule{15-16}\cmidrule{18-19}
Item                                     & \multicolumn{1}{l}{Factor 1}              & \multicolumn{1}{l}{Factor 2}              & \multicolumn{1}{l}{Uniqu.}               &                      & \multicolumn{1}{l}{Factor 1}             & \multicolumn{1}{l}{Uniqu.}               &                      & \multicolumn{1}{l}{Factor 1}             & \multicolumn{1}{l}{Uniqu.}               &                      & \multicolumn{1}{l}{Factor 1}             & \multicolumn{1}{l}{Uniqu.}               &                      & \multicolumn{1}{l}{Factor 1}              & \multicolumn{1}{l}{Uniqu.}               &                      & \multicolumn{1}{l}{Factor 1}              & \multicolumn{1}{l}{Uniqu.}                \\ 
\midrule
T1\_PF1\_ExerciseComp                    & 0.54                                      & \textcolor[rgb]{0.900,0.900,0.900}{0.10}  & 0.64                                     &                      & \textcolor[rgb]{0.900,0.900,0.900}{0.17} & \textcolor[rgb]{0.900,0.900,0.900}{0.97} &                      & 0.54                                     & 0.71                                     &                      & 0.50                                     & 0.75                                     &                      & 0.71                                      & 0.50                                     &                      & 0.66                                      & 0.57                                      \\
T1\_PF2\_LearningComp                    & 0.57                                      & \textcolor[rgb]{0.900,0.900,0.900}{0.10}  & 0.60                                     &                      & 0.48                                     & 0.77                                     &                      & 0.52                                     & 0.73                                     &                      & 0.60                                     & 0.64                                     &                      & 0.73                                      & 0.47                                     &                      & 0.54                                      & 0.70                                      \\
T1\_PF3\_CodingFun                       & 0.69                                      & \textcolor[rgb]{0.900,0.900,0.900}{0.11}  & 0.43                                     &                      & 0.54                                     & 0.71                                     &                      & 0.74                                     & 0.45                                     &                      & 0.67                                     & 0.55                                     &                      & 0.79                                      & 0.38                                     &                      & 0.69                                      & 0.52                                      \\
T1\_RE1\_CSInterest                      & 0.97                                      & \textcolor[rgb]{0.900,0.900,0.900}{-0.18} & 0.20                                     &                      & 0.74                                     & 0.46                                     &                      & 0.84                                     & 0.30                                     &                      & 0.61                                     & 0.63                                     &                      & 0.87                                      & 0.24                                     &                      & 0.79                                      & 0.38                                      \\
T1\_RE2\_CodingEveryday                  & \textcolor[rgb]{0.900,0.900,0.900}{-0.07} & 0.67                                      & 0.59                                     &                      & \textcolor[rgb]{0.900,0.900,0.900}{0.12} & \textcolor[rgb]{0.900,0.900,0.900}{0.99} &                      & \textcolor[rgb]{0.900,0.900,0.900}{0.29} & \textcolor[rgb]{0.900,0.900,0.900}{0.92} &                      & 0.44                                     & 0.81                                     &                      & 0.33                                      & 0.89                                     &                      & \textcolor[rgb]{0.900,0.900,0.900}{0.16}  & \textcolor[rgb]{0.900,0.900,0.900}{0.98}  \\
T1\_RE3\_CareerRelevance                 & 0.22                                      & 0.64                                      & 0.40                                     &                      & 0.76                                     & 0.42                                     &                      & 0.56                                     & 0.69                                     &                      & 0.69                                     & 0.52                                     &                      & 0.58                                      & 0.66                                     &                      & \textcolor[rgb]{0.900,0.900,0.900}{0.28}  & \textcolor[rgb]{0.900,0.900,0.900}{0.92}  \\
T1\_RE4\_CSJobsBoring               & 0.56                                      & \textcolor[rgb]{0.900,0.900,0.900}{0.17}  & 0.56                                     &                      & 0.75                                     & 0.43                                     &                      & 0.63                                     & 0.61                                     &                      & 0.75                                     & 0.44                                     &                      & 0.69                                      & 0.52                                     &                      & 0.63                                      & 0.61                                      \\
T1\_RE5\_CSTopicsInteresting             & 0.42                                      & 0.34                                      & 0.56                                     &                      & 0.75                                     & 0.44                                     &                      & 0.56                                     & 0.69                                     &                      & 0.75                                     & 0.44                                     &                      & 0.67                                      & 0.55                                     &                      & 0.37                                      & 0.86                                      \\
T1\_FI1\_NoCoding                   & 0.51                                      & \textcolor[rgb]{0.900,0.900,0.900}{0.15}  & 0.65                                     &                      & \multicolumn{2}{r}{not included}                                                    &                      & 0.71                                     & 0.50                                     &                      & \textcolor[rgb]{0.900,0.900,0.900}{0.23} & \textcolor[rgb]{0.900,0.900,0.900}{0.95} &                      & 0.58                                      & 0.66                                     &                      & 0.52                                      & 0.73                                      \\
T1\_FI2\_LearningMore                    & 0.66                                      & \textcolor[rgb]{0.900,0.900,0.900}{0.14}  & 0.44                                     &                      & 0.60                                     & 0.64                                     &                      & 0.75                                     & 0.44                                     &                      & 0.77                                     & 0.41                                     &                      & 0.73                                      & 0.46                                     &                      & 0.72                                      & 0.49                                      \\
T1\_FI3\_Career                          & 0.71                                      & \textcolor[rgb]{0.900,0.900,0.900}{0.13}  & 0.38                                     &                      & 0.80                                     & 0.36                                     &                      & 0.81                                     & 0.34                                     &                      & 0.68                                     & 0.53                                     &                      & 0.74                                      & 0.45                                     &                      & 0.69                                      & 0.52                                      \\
T1\_SE\_SelfEfficacy                     & 0.70                                      & \textcolor[rgb]{0.900,0.900,0.900}{-0.05} & 0.54                                     &                      & 0.66                                     & 0.57                                     &                      & 0.66                                     & 0.57                                     &                      & 0.54                                     & 0.71                                     &                      & 0.66                                      & 0.57                                     &                      & 0.42                                      & 0.83                                      \\
T1\_ST\_Stereotypes                      & \textcolor[rgb]{0.900,0.900,0.900}{0.12}  & \textcolor[rgb]{0.900,0.900,0.900}{0.25}  & \textcolor[rgb]{0.900,0.900,0.900}{0.89} &                      & -0.31                                    & 0.91                                     &                      & 0.39                                     & 0.85                                     &                      & 0.79                                     & 0.37                                     &                      & \textcolor[rgb]{0.900,0.900,0.900}{0.28}  & \textcolor[rgb]{0.900,0.900,0.900}{0.92} &                      & \textcolor[rgb]{0.900,0.900,0.900}{0.19}  & \textcolor[rgb]{0.900,0.900,0.900}{0.96}  \\
T1\_PE2\_CSJobsKnowing                   & 0.47                                      & 0.77                                      & \multicolumn{1}{l}{}                     &                      & \multicolumn{2}{r}{not included}                                                    &                      & 0.53                                     & 0.72                                     &                      & \textcolor[rgb]{0.900,0.900,0.900}{0.23} & \textcolor[rgb]{0.900,0.900,0.900}{0.95} &                      & 0.49                                      & 0.76                                     &                      & \multicolumn{2}{r}{not included}                                                      \\
T1\_PE4\_CSOnlyCoding                    & \textcolor[rgb]{0.900,0.900,0.900}{0.06}  & \textcolor[rgb]{0.900,0.900,0.900}{0.09}  & \textcolor[rgb]{0.900,0.900,0.900}{0.98} &                      & 0.67                                     & 0.55                                     &                      & \textcolor[rgb]{0.900,0.900,0.900}{0.03} & \textcolor[rgb]{0.900,0.900,0.900}{1.00} &                      & 0.34                                     & 0.88                                     &                      & \textcolor[rgb]{0.900,0.900,0.900}{-0.12} & \textcolor[rgb]{0.900,0.900,0.900}{0.99} &                      & \textcolor[rgb]{0.900,0.900,0.900}{-0.08} & \textcolor[rgb]{0.900,0.900,0.900}{0.99}  \\
T1\_PE5\_CSEverywhere                    & \textcolor[rgb]{0.900,0.900,0.900}{0.04}  & 0.38                                      & 0.84                                     &                      & 0.81                                     & 0.35                                     &                      & \textcolor[rgb]{0.900,0.900,0.900}{0.17} & \textcolor[rgb]{0.900,0.900,0.900}{0.97} &                      & 0.34                                     & 0.88                                     &                      & 0.42                                      & 0.82                                     &                      & \textcolor[rgb]{0.900,0.900,0.900}{0.08}  & \textcolor[rgb]{0.900,0.900,0.900}{0.99}  \\ 
\midrule
Factor variance explained in \%          & 30.7                                      & 10.4                                      &                                          & \multicolumn{1}{r}{} & 38.8                                     &                                          & \multicolumn{1}{r}{} & 34.5                                     &                                          & \multicolumn{1}{r}{} & 34.7                                     &                                          & \multicolumn{1}{r}{} & 38.6                                      &                                          & \multicolumn{1}{r}{} & 26.4                                      &                                           \\
Mean of uniqueness                     &                                           &                                           & 0.58                                     & \multicolumn{1}{r}{} &                                          & 0.61                                     & \multicolumn{1}{r}{} &                                          & 0.66                                     & \multicolumn{1}{r}{} &                                          & 0.65                                     & \multicolumn{1}{r}{} &                                           & 0.62                                     & \multicolumn{1}{r}{} &                                           & 0.74                                      \\ 
\midrule
RMSEA                                    & 0.08                                      & \multicolumn{1}{l}{}                      & \multicolumn{1}{l}{}                     &                      & 0.11                                     & \multicolumn{1}{l}{}                     &                      & 0.09                                     & \multicolumn{1}{l}{}                     &                      & 0.06                                     & \multicolumn{1}{l}{}                     &                      & 0.12                                      & \multicolumn{1}{l}{}                     &                      & 0.08                                      & \multicolumn{1}{l}{}                      \\
TLI                                      & 0.88                                      & \multicolumn{1}{l}{}                      & \multicolumn{1}{l}{}                     &                      & 0.55                                     & \multicolumn{1}{l}{}                     &                      & 0.82                                     & \multicolumn{1}{l}{}                     &                      & 0.84                                     & \multicolumn{1}{l}{}                     &                      & 0.74                                      & \multicolumn{1}{l}{}                     &                      & 0.75                                      & \multicolumn{1}{l}{}                      \\
Model fit $\chi^{2}$ (df); p                     & \multicolumn{2}{c}{190.02 (89); <.001}                                                 &                                          & \multicolumn{1}{r}{} & \multicolumn{2}{c}{96.6 (77); .065}                                                 & \multicolumn{1}{r}{} & \multicolumn{2}{c}{160.5 (104); <.001}                                               & \multicolumn{1}{r}{} & \multicolumn{2}{c}{121.6 (104); .114}                                               & \multicolumn{1}{r}{} & \multicolumn{2}{c}{194.1 (104); <.001}                                                & \multicolumn{1}{r}{} & \multicolumn{2}{c}{118.2 (90); .025}                                                  \\
Bartlett's Test of Sphericity $\chi^{2}$ (df); p & \multicolumn{2}{c}{1256.87 (120); <.001}                                               &                                          & \multicolumn{1}{r}{} & \multicolumn{2}{c}{153 (91); <.001}                                                  & \multicolumn{1}{r}{} & \multicolumn{2}{c}{494.8 (120); <.001}                                               & \multicolumn{1}{r}{} & \multicolumn{2}{c}{255.6 (120); <.001}                                               & \multicolumn{1}{r}{} & \multicolumn{2}{c}{530.0 (120); <.001}                                                & \multicolumn{1}{r}{} & \multicolumn{2}{c}{242.8 (105); <.001}                                                 \\
KMO overall                              & 0.89                                      & \multicolumn{1}{l}{}                      & \multicolumn{1}{l}{}                     &                      & 0.42                                     & \multicolumn{1}{l}{}                     &                      & 0.81                                     & \multicolumn{1}{l}{}                     &                      & 0.71                                     & \multicolumn{1}{l}{}                     &                      & 0.83                                      & \multicolumn{1}{l}{}                     &                      & 0.68                                      & \multicolumn{1}{l}{}                      \\
\bottomrule
\end{tabular}
}
\end{table*}
\end{landscape}

\subsection{Effect of Gender and Age on CS Interest and Perception}

We conducted a two-way ANOVA on pre-test data to examine the effects of gender and age on students' interest in CS and their perceptions of the subject prior to participating in an educational intervention.

The analysis revealed no significant interaction between age and gender, indicating that the effects of these variables on CS interest and perception are independent. However, main effects were observed (see Table \ref{tab:anova}): age had a statistically significant effect on all three dimensions of interest with large effect sizes observed ($\eta^2$>0.22) as well as on self-efficacy with a medium effect size, whereas gender significantly affected future intents, positive feelings, and self-efficacy but with lower effect sizes (ranging from low to medium) and insufficient power to conclusively attribute effects to gender differences. This suggests that age is a more determinant factor in shaping students' attitudes towards CS. The differences in stereotype perception also differed significantly for age, but the effect size and statistical power are poor.

\begin{table}
\centering
    \caption{Mean values for girls and boys in the pre-test for the three dimensions of interest (PF-positive feelings, RE-relevance, FI-future intents), SE-self-efficacy, and ST-stereotype with test statistics from two-way ANOVA (F ratio), partial eta squared for effect sizes ($\eta^{2}$), statistical significance (p), and observed power.}
    \label{tab:anova}
\begin{adjustbox}{width=\linewidth}

\centering
\scriptsize 
\begin{tabular}{lrrrrrrrrrrrrrrrrr}
\toprule
\multicolumn{1}{r}{} & \multicolumn{3}{l}{Girls} &  & \multicolumn{3}{l}{Boys} &  & \multicolumn{4}{l}{Gender}     &  & \multicolumn{4}{l}{Age Group (AG)}  \\ 
\cmidrule(lr){2-4}\cmidrule(lr){6-8}\cmidrule(lr){10-13}\cmidrule(lr){15-18}
\multicolumn{1}{r}{} & AG1  & AG2  & AG3         &  & AG1  & AG2  & AG3        &  & F ratio & $\eta^{2}$   & p     & power &  & F ratio & $\eta^{2}$   & p    & power       \\ 
\midrule
PF                   & 3.93 & 4.09 & 2.50        &  & 4.30 & 4.45 & 3.15       &  & 26.30    & 0.06 & $<$.001  & 0.96  &  & 92.55   & 0.32 & .001 & 1.00        \\
RE                   & 3.75 & 3.93 & 2.89        &  & 3.94 & 4.17 & 3.22       &  & 10.65   & 0.03 & .001 & 0.48  &  & 56.22   & 0.22 & $<$.001 & 1.00        \\
FI                   & 3.26 & 3.59 & 1.92        &  & 3.55 & 4.02 & 2.37       &  & 13.95   & 0.03 & $<$.001  & 0.66  &  & 88.34   & 0.31 & $<$.001 & 1.00        \\
SE                   & 3.49 & 3.51 & 2.67        &  & 3.94 & 4.03 & 3.22       &  & 19.64   & 0.05 & $<$.001  & 0.87  &  & 18.64   & 0.09 & $<$.001 & 0.99        \\
ST                   & 4.53 & 4.78 & 4.25        &  & 4.19 & 4.61 & 4.65       &  & 0.13    & 0.00 & .716 & 0.00  &  & 7.32    & 0.04 & $<$.001 & 0.58        \\
\bottomrule
\end{tabular}

\end{adjustbox}    
\end{table}

Further investigation using one-way ANOVA on the individual items associated with the three interest dimensions showed that age group significantly influenced all items (p < .001), as corroborated by Welch's test results. Effect sizes were large across all items ($\eta^{2}>$0.12). The highest effect sizes are observed for \emph{T1\_PF3\_CodingFun} ($\eta^{2}$=0.38), \emph{T1\_FI2\_LearningMore} ($\eta^{2}$=0.37), \emph{T1\_RE1\_CSInterest} ($\eta^{2}$=0.36), \emph{T1\_PF1\_ExerciseComp} ($\eta^{2}$=0.35), and \emph{T1\_FI3\_Career} ($\eta^{2}$=0.34). These findings indicate significant age-related differences in associations with activities such as computer work and coding enjoyment, overall interest in CS, and future career intentions within the field. 
The smallest effect sizes were recorded for \emph{T1\_RE3\_CareerRelevance} ($\eta^{2}$=0.12) and \emph{T1\_RE5\_CSTopicsInteresting} ($\eta^{2}$=0.13). Statistical power was 1.00 for tests.

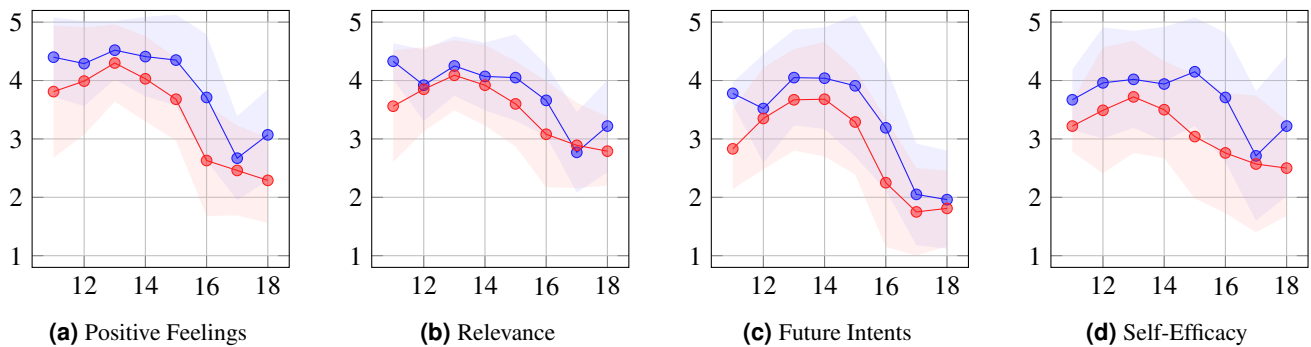
\begin{figure}
    \centering
    \begin{subfigure}[b]{0.23\textwidth}
         \centering
\begin{tikzpicture}
\begin{axis}[
height=5cm,
width=5cm,
ymin=0.8,ymax=5.2,
grid=major,
]


\addplot[blue!90,fill opacity=0.5,mark=*] coordinates {
(11,4.40)
(12,4.29)
(13,4.52)
(14,4.41)
(15,4.35)
(16,3.71)
(17,2.67)
(18,3.07)
};

\addplot[name path=upper,draw=none] coordinates {
(11,4.40+0.68)
(12,4.29+0.73)
(13,4.52+0.50)
(14,4.41+0.68)
(15,4.35+0.78)
(16,3.71+1.08)
(17,2.67+0.72)
(18,3.07+0.78)
};

\addplot[name path=lower,draw=none] coordinates {
(11,4.40-0.68)
(12,4.29-0.73)
(13,4.52-0.50)
(14,4.41-0.68)
(15,4.35-0.78)
(16,3.71-1.08)
(17,2.67-0.72)
(18,3.07-0.78)
};

\addplot[blue!30, fill opacity=0.2] fill between[of=upper and lower];

\addplot[red!90,fill opacity=0.5,mark=*] coordinates {
(11,3.81)
(12,3.99)
(13,4.30)
(14,4.03)
(15,3.68)
(16,2.63)
(17,2.46)
(18,2.29)
};

\addplot[name path=upper,draw=none] coordinates {
(11,3.81+1.13)
(12,3.99+0.92)
(13,4.30+0.66)
(14,4.03+0.73)
(15,3.68+0.71)
(16,2.63+0.95)
(17,2.46+0.77)
(18,2.29+0.73)
};

\addplot[name path=lower,draw=none] coordinates {
(11,3.81-1.13)
(12,3.99-0.92)
(13,4.30-0.66)
(14,4.03-0.73)
(15,3.68-0.71)
(16,2.63-0.95)
(17,2.46-0.77)
(18,2.29-0.73)
};
\addplot[red!30, fill opacity=0.2] fill between[of=upper and lower];

\end{axis}
\end{tikzpicture}
         \caption{Positive Feelings}
         \label{subfig:pf}
     \end{subfigure}
     \hfill
     \begin{subfigure}[b]{0.23\textwidth}
         \centering
\begin{tikzpicture}
\begin{axis}[
height=5cm,
width=5cm,
grid=major,
ymin=0.8,ymax=5.2,
]


\addplot[blue!90,fill opacity=0.5,mark=*] coordinates {
(11,4.33)
(12,3.92)
(13,4.25)
(14,4.07)
(15,4.05)
(16,3.66)
(17,2.77)
(18,3.22)
};

\addplot[name path=upper,draw=none] coordinates {
(11,4.33+0.31)
(12,3.92+0.61)
(13,4.25+0.51)
(14,4.07+0.58)
(15,4.05+0.74)
(16,3.66+0.61)
(17,2.77+0.69)
(18,3.22+0.79)
};

\addplot[name path=lower,draw=none] coordinates {
(11,4.33-0.31)
(12,3.92-0.61)
(13,4.25-0.51)
(14,4.07-0.58)
(15,4.05-0.74)
(16,3.66-0.61)
(17,2.77-0.69)
(18,3.22-0.79)
};

\addplot[blue!30, fill opacity=0.2] fill between[of=upper and lower];

\addplot[red!90,fill opacity=0.5,mark=*] coordinates {
(11,3.56)
(12,3.85)
(13,4.09)
(14,3.92)
(15,3.60)
(16,3.08)
(17,2.89)
(18,2.79)
};

\addplot[name path=upper,draw=none] coordinates {
(11,3.56+0.96)
(12,3.85+0.72)
(13,4.09+0.60)
(14,3.92+0.69)
(15,3.60+0.73)
(16,3.08+0.91)
(17,2.89+0.73)
(18,2.79+0.59)
};

\addplot[name path=lower,draw=none] coordinates {
(11,3.56-0.96)
(12,3.85-0.72)
(13,4.09-0.60)
(14,3.92-0.69)
(15,3.60-0.73)
(16,3.08-0.91)
(17,2.89-0.73)
(18,2.79-0.59)
};
\addplot[red!30, fill opacity=0.2] fill between[of=upper and lower];

\end{axis}
\end{tikzpicture}
         \caption{Relevance}
         \label{subfif:re}
     \end{subfigure}
     \hfill
     \begin{subfigure}[b]{0.23\textwidth}
         \centering
\begin{tikzpicture}
\begin{axis}[
height=5cm,
width=5cm,
grid=major,
ymin=0.8,ymax=5.2,
]


\addplot[blue!90,fill opacity=0.5,mark=*] coordinates {
(11,3.78)
(12,3.52)
(13,4.05)
(14,4.04)
(15,3.91)
(16,3.19)
(17,2.05)
(18,1.96)
};

\addplot[name path=upper,draw=none] coordinates {
(11,3.78+0.19)
(12,3.52+0.93)
(13,4.05+0.82)
(14,4.04+0.87)
(15,3.91+1.21)
(16,3.19+1.07)
(17,2.05+0.87)
(18,1.96+0.84)
};

\addplot[name path=lower,draw=none] coordinates {
(11,3.78-0.19)
(12,3.52-0.93)
(13,4.05-0.82)
(14,4.04-0.87)
(15,3.91-1.21)
(16,3.19-1.07)
(17,2.05-0.87)
(18,1.96-0.84)
};

\addplot[blue!30, fill opacity=0.2] fill between[of=upper and lower];






\addplot[red!90,fill opacity=0.5,mark=*] coordinates {
(11,2.83)
(12,3.35)
(13,3.67)
(14,3.68)
(15,3.29)
(16,2.25)
(17,1.75)
(18,1.81)
};

\addplot[name path=upper,draw=none] coordinates {
(11,2.83+0.69)
(12,3.35+0.87)
(13,3.67+0.87)
(14,3.68+0.98)
(15,3.29+0.90)
(16,2.25+1.10)
(17,1.75+0.75)
(18,1.81+0.65)
};

\addplot[name path=lower,draw=none] coordinates {
(11,2.83-0.69)
(12,3.35-0.87)
(13,3.67-0.87)
(14,3.68-0.98)
(15,3.29-0.90)
(16,2.25-1.10)
(17,1.75-0.75)
(18,1.81-0.65)
};
\addplot[red!30, fill opacity=0.2] fill between[of=upper and lower];




\end{axis}
\end{tikzpicture}
         \caption{Future Intents}
         \label{subfig:FI}
     \end{subfigure}
     \hfill
     \begin{subfigure}[b]{0.23\textwidth}
         \centering
\begin{tikzpicture}
\begin{axis}[
height=5cm,
width=5cm,
grid=major,
ymin=0.8,ymax=5.2,
]


\addplot[blue!90,fill opacity=0.5,mark=*] coordinates {
(11,3.67)
(12,3.96)
(13,4.02)
(14,3.94)
(15,4.15)
(16,3.71)
(17,2.71)
(18,3.22)
};

\addplot[name path=upper,draw=none] coordinates {
(11,3.67+0.52)
(12,3.96+0.95)
(13,4.02+0.83)
(14,3.94+0.98)
(15,4.15+0.93)
(16,3.71+1.11)
(17,2.71+1.11)
(18,3.22+1.20)
};

\addplot[name path=lower,draw=none] coordinates {
(11,3.67-0.52)
(12,3.96-0.95)
(13,4.02-0.83)
(14,3.94-0.98)
(15,4.15-0.93)
(16,3.71-1.11)
(17,2.71-1.11)
(18,3.22-1.20)
};

\addplot[blue!30, fill opacity=0.2] fill between[of=upper and lower];

\addplot[red!90,fill opacity=0.5,mark=*] coordinates {
(11,3.22)
(12,3.49)
(13,3.72)
(14,3.50)
(15,3.04)
(16,2.76)
(17,2.57)
(18,2.50)
};

\addplot[name path=upper,draw=none] coordinates {
(11,3.22+0.44)
(12,3.49+1.08)
(13,3.72+0.96)
(14,3.50+0.83)
(15,3.04+1.06)
(16,2.76+1.03)
(17,2.57+1.17)
(18,2.50+0.82)
};

\addplot[name path=lower,draw=none] coordinates {
(11,3.22-0.44)
(12,3.49-1.08)
(13,3.72-0.96)
(14,3.50-0.83)
(15,3.04-1.06)
(16,2.76-1.03)
(17,2.57-1.17)
(18,2.50-0.82)
};
\addplot[red!30, fill opacity=0.2] fill between[of=upper and lower];

\end{axis}
\end{tikzpicture}
         \caption{Self-Efficacy}
         \label{subfig:SE}
     \end{subfigure}
    \caption{Plots of mean values of girls' (red) and boys' (blue) responses in the pre-test (T1) to the self-efficacy item and summarized for items related to one of the three dimensions of interest. The colored area illustrates standard derivation.}
    \label{fig:age-line-charts}
\end{figure}

Pairwise comparisons of age groups and gender revealed notable trends in students' interest and self-efficacy in CS (see Figure \ref{fig:age-line-charts}). Significant differences (p < .001) emerged between AG3 (16 to 19 years) and the younger cohorts, AG1 (10 to 12 years) and AG2 (13 to 15 years), whereas differences between AG1 and AG2 were not statistically significant. Notably, a considerable decline in students' future intentions in CS was observed from AG2 to AG3, with mean differences of -1.76 from AG3 to AG2 and -1.49 from AG3 to AG1, in contrast to an initial increase from AG1 to AG2 (mean difference: +0.71). Furthermore, the plots (Figure \ref{fig:age-line-charts}) show an overall upward trend in interest through age 13, followed by a decline, particularly pronounced at ages 15 and 16. Although the interest trajectories for both genders run parallel, females consistently display slightly lower interest than males. In terms of self-efficacy, a gender-specific pattern emerges: girls show a notable earlier decrease around age 13, whereas boys show a decrease about two years later, at age 15.

\subsection{Effect of Age on Interest and Perception Change}
This section explores the interplay between age and the enthusiasm potential of interventions, measured by changes in students' attitudes toward CS, by comparing pre-test and post-test means. 

While only changes in the item \emph{TX\_FI1\_NoCoding} did statistically significantly differ between age groups (p$<$.001), when examining the average mean differences across all items, distinct trends emerge among the age groups. Specifically, AG3 exhibited the largest change, with an average difference of dif=+0.20, 43\% greater than for AG1 (dif=+0.14) and 220\% greater than for AG2 (dif=+0.09). This pattern indicates that interventions have greater potential for enthusiasm among older students than younger students, as evidenced by more pronounced effects on their attitudes and perceptions. However, this difference might also be influenced by the lower baseline mean score for AG3 (M=2.99) compared with AG1 (M=3.68) and AG2 (M=3.94), which allows for larger shifts.

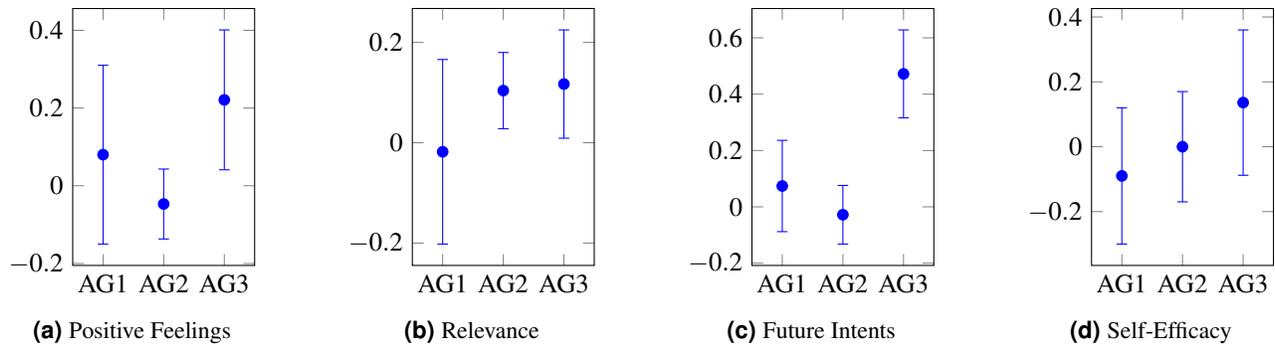
\begin{figure}
    \begin{subfigure}[b]{0.23\textwidth}
         \centering
         \begin{tikzpicture}
\begin{axis}[
height=5cm,
xtick={1, 2, 3},
xticklabels={AG1, AG2, AG3},
width=4cm,
xmin=0.5, xmax=3.5,
]


\addplot[blue, mark=*, only marks] plot[error bars/.cd,
y dir=both,y explicit]
coordinates {
(1,0.080) +- (0,2*0.115) 
(2,-0.047) +- (0,2*0.045) 
(3,0.221) +- (0,2*0.090) 
};
\end{axis}
\end{tikzpicture}
         \caption{Positive Feelings}
         \label{subfig:dif-PF}
     \end{subfigure}
     \hfill
     \begin{subfigure}[b]{0.23\textwidth}
         \centering
         \begin{tikzpicture}
\begin{axis}[
height=5cm,
xtick={1, 2, 3},
xticklabels={AG1, AG2, AG3},
width=4cm,
xmin=0.5, xmax=3.5,
]


\addplot[blue, mark=*, only marks] plot[error bars/.cd,
y dir=both,y explicit]
coordinates {
(1,-0.018) +- (0,2*0.092) 
(2,0.104) +- (0,2*0.038) 
(3,0.117) +- (0,2*0.054) 
};

\end{axis}
\end{tikzpicture}
         \caption{Relevance}
         \label{subfig:dif-RE}
     \end{subfigure}
     \hfill
     \begin{subfigure}[b]{0.23\textwidth}
         \centering
         \begin{tikzpicture}
\begin{axis}[
height=5cm,
xtick={1, 2, 3},
xticklabels={AG1, AG2, AG3},
width=4cm,
xmin=0.5, xmax=3.5,
]


\addplot[blue, mark=*, only marks] plot[error bars/.cd,
y dir=both,y explicit]
coordinates {
(1,0.074) +- (0,2*0.081) 
(2,-0.028) +- (0,2*0.052) 
(3,0.472) +- (0,2*0.078) 
};

\end{axis}
\end{tikzpicture}
         \caption{Future Intents}
         \label{subfig:dif-FI}
     \end{subfigure}
     \hfill
     \begin{subfigure}[b]{0.23\textwidth}
         \centering
         \begin{tikzpicture}
\begin{axis}[
height=5cm,
xtick={1, 2, 3},
xticklabels={AG1, AG2, AG3},
width=4cm,
xmin=0.5, xmax=3.5,
]


\addplot[blue, mark=*, only marks] plot[error bars/.cd,
y dir=both,y explicit]
coordinates {
(1,-0.090) +- (0,2*0.105) 
(2,0.000) +- (0,2*0.085) 
(3,0.136) +- (0,2*0.112) 
};

\end{axis}
\end{tikzpicture}
         \caption{Self-Efficacy}
         \label{subfig:dif-SE}
     \end{subfigure}
    \caption{Mean differences between pre-test and post-test with 95\%-CI (y-axis) for items related to one of the three dimensions of interest as well as self-efficacy by age groups (AGs, x-axis).}
    \label{fig:dif_plot}
\end{figure}

Figure \ref{fig:dif_plot} shows the mean differences summarized for the three dimensions of interest as well as for self-efficacy by age groups. For AG1 and AG2, effects on student interest and self-efficacy were generally neutral, except for changes in perceived relevance in AG2. In contrast, AG3 displayed the most substantial changes across all categories, notably in future intentions in CS. The willingness of AG3 students to engage in CS increased significantly from pre-test to post-test. 
The greatest changes were observed for the item \emph{TX\_FI1\_NoCoding}, which showed an average increase of dif=+0.79 from pre-test (M=3.81) to post-test (M=3.08), reflecting a heightened interest in coding activities among older students through course participation. Similarly, other items related to future intentions in CS for AG3 indicated notable average increases of dif=+0.35 for \emph{TX\_FI2\_LearningMore} from M=2.39 to M=2.76 and dif=+0.33 for \emph{TX\_FI3\_Career} from M=1.75 to M=2.11, although these were not statistically significant across different age groups (p=.002 and p=.070, respectively).

Moreover, self-efficacy, although not statistically significant across age groups (p=.338), showed the greatest increase among AG3 students, whereas younger students, particularly those in AG1, partially showed downward trends in self-efficacy scores.

\subsection{Effect of Age on Course-related Attitudes}
This section analyzes student attitudes toward course activities, focusing on the post-test-only assessment items (labeled \emph{T2C}).

Analysis of the mean values across all \emph{T2C} items reveals that the courses were better received by the younger age groups, AG1 (M${avg}$=3.68) and AG2 (M${avg}$=3.86), compared to the older students in AG3 (M$_{avg}$=3.32). This difference was statistically significant for six out of the ten items.

Significant age-related effects were identified, with medium to large effect sizes noted for \emph{T2C\_PF2\_Curiosity} ($\eta^{2}$=0.09) and \emph{T2C\_FI3\_Talk} ($\eta^{2}$=0.09), indicating a heightened curiosity about CS among younger students and an increased likelihood of discussing their course experiences externally. Medium effect sizes were observed for \emph{T2C\_PF1\_Fun} ($\eta^{2}$=0.07) and \emph{T2C\_FI1\_Repeat} ($\eta^{2}$=0.06), suggesting younger students found the course activities more enjoyable and showed a higher interest in enrolling in similar courses in the future, all of which were statistically significant (p$<$.001).

The item \emph{T2C\_IIN\_InterestIncrease} showed particularly notable differences, with AG3 students reporting significantly lower interest levels (M=2.76) compared to AG1 (M=3.47) and AG2 (M=3.49) students (F(2)=7.52, p$<$.001, $\eta^{2}$=0.05). Although AG3 experienced the most significant changes through the intervention, they reported the lowest direct impact on their interest in CS.

\section{Discussion}

In general, a girl's age has always been considered important to their interest in CS, but studies have usually been limited to stating that with girls' age, interest in CS is declining \cite{gursch2022coding, palma2001women} or where investigating age in adults \cite{ruthotto2020lurking}. 
However, as a determinant of variation in this interest and its factors, it has rarely been investigated empirically. We are not aware of any studies that examine age as a primary determinant of its influence and its interactions with other variables. Several studies have found that early engagement in CS among girls is very important \cite{kekelis2005hurdles}, but they do not examine the issue in depth. Our research brings a detailed picture of influence and interaction in different age groups. 
Exploratory factor analysis revealed significant variation in factor loadings across age groups, indicating an evolution in the nature of the latent factors underlying interests and perceptions of CS. These findings align with occupational, career, and study theories, which posit a gradual development of this decision during adolescence \cite{eccles1994understanding, super1980life}. Our empirical results suggest a dynamic framework in which girls' core interests remain stable while priorities shift as they age \cite{dasgupta2014girls}, consistent with similar findings for stereotypes \cite{master2021gender}. Our ANOVA analysis further supported these findings, showing that both gender and age significantly influence students’ interest and self-efficacy in CS.
Our findings extend existing knowledge, which has investigated only age \cite{mensah2019computer} or only efficacy in adults, finding that efficacy declines in older adults \cite{chu2010family}.

Notably, age emerged as the predominant factor, evidenced by large effect sizes, whereas gender showed only small to medium effects. The trend indicates an increase in interest and self-efficacy from ages 14 to 15, followed by a decline. Specifically, the mean values for AG3 were significantly lower than those for AG1 and AG2, highlighting a marked decrease in CS interest across all items surveyed. 
This distinction is crucial, as it suggests that educational strategies should be more finely tuned to age-related changes rather than broadly addressing gender disparities. 

The decline in interest and self-efficacy observed around ages 14 to 15 raises questions about the nature of CS education and its alignment with students' developmental needs. Contrary to our initial expectation that interventions might be less effective for older students due to their purportedly waning interest, our study found that these interventions were most effective for the older age group (AG3), even though they were initially designed for a younger audience. This paradox underscores the complexity of measuring and interpreting interest and self-efficacy, suggesting that even when students' self-reported interest does not increase, their engagement with and attitudes towards CS can improve significantly through well-designed interventions.

Our observations highlight the nuanced relationship between situational and individual interest. It is possible that the interventions tapped into older students' situational interest, creating a temporary spike in enthusiasm that did not translate into a long-term increase in individual interest. This phenomenon suggests that interventions might need to be repeated or sustained over a longer period to have a lasting impact on students' overall interest in CS. This underscores the mutual relationship between situational and individual interest, where regular exposure to situational interest plays a pivotal role in cultivating individual interest \cite{rotgans2017interest}. Our analysis further suggests that this dynamic is especially pertinent during later adolescent stages.

Reflecting on the mitigating effect of direct computer experience on negative perceptions, particularly among older and female students, emphasizes the critical role of consistent exposure and familiarity in shaping students' attitudes towards CS. Our findings align with those of Comber et al. \cite{comber1997effects} regarding gender differences in CS interest, showing a general decrease with age for both genders, unlike the gender-specific trends observed in Comber's study. This broader decline underscores the impact of gender stereotypes, as outlined by Master et al. \cite{master2016computing}, which not only influence girls' motivation toward CS but also mediate their sense of belonging and interest in the field. The mitigating effect of direct computer experience on negative perceptions, particularly among older and female students, as highlighted by Comber and colleagues \cite{comber1997effects, colley2003age}, emphasizes the importance of exposure and familiarity in fostering positive attitudes towards CS.

The decline in interest among older students, especially girls, may also reflect broader societal and psychological challenges. 
Cavanagh \cite{cavanagh2007puberty} suggests strategies to minimize the social and psychological challenges faced by girls during early puberty, which can disrupt their academic pursuits. This includes limiting their exposure to older adolescents, especially boys, and supporting advocacy for educational settings that protect young girls from negative influences during critical developmental periods. Adolescence is a period of profound psychological change, during which students undergo shifts in cognitive and emotional development. For girls, early puberty brings additional challenges, including heightened self-consciousness and increased susceptibility to peer and societal pressures. These changes can influence academic interests, leading to reduced engagement with subjects perceived as incongruent with their developing identity or societal expectations. Here are the social and psychological challenges arising from the educational environment, which are significant factors in sustaining or diminishing students' interest in CS.

Moreover, the observation that older students perceived the overall course less favorably than younger students, yet still experienced an improvement in their attitude towards CS, underscores the effectiveness of interventions tailored to the specific needs of different age groups. This indicates that while younger children may approach CS with novelty and enthusiasm, older students' engagement is driven more by cognitive than affective factors \cite{fandakova2021states}, leading to a more critical perspective and a nuanced understanding of the field, necessitating interventions that address their unique interests and concerns. Factors such as societal stereotypes, environmental conditions, and evolving peer and societal influences further contribute to the complexity of changing perceptions of CS as students age.

Taken together, our findings illustrate that enthusiasm—conceptualized as a short-term activation of affective and value-related components of interest—varies meaningfully across adolescence and remains responsive to brief CS learning activities. These insights underscore the importance of designing age-sensitive CS interventions and the value of using intervention-sensitive measures to capture attitudinal dynamics that might otherwise remain undetected.

\section{Limitations}
Our study, while contributing valuable insights into the dynamics of CS interest among students, is subject to certain limitations that warrant consideration. 
First, while significant emphasis was placed on ensuring precision and accuracy in the questionnaire's development, its sensitivity for measuring longitudinal shifts in CS attitudes remains an area for iterative refinement. Although the CFA and reliability results confirm the instrument's capacity to measure the intended latent constructs, the fit indices suggest that further optimization of specific items could enhance its psychometric robustness in future research.

Second, the distribution of sample sizes across age groups posed another challenge, introducing variability that may affect the generalizability of our findings. Additionally, achieving a deeper understanding of how CS interest develops at specific stages of adolescence necessitates larger sample sizes. In the future, we aim to collect larger, more balanced samples, enabling more nuanced analysis and insights into subtle trends and patterns that our current study may not fully capture.

Concerning our statistical approach, the data's deviation from normal distribution prompted the use of ANOVA despite its assumptions. However, ANOVA's robustness to certain violations of normality \cite{glass1972consequences, harwell1992summarizing}, coupled with our use of Welch's test for one-way ANOVA \cite{kim2014analysis}, lends credibility to our statistical findings.

\section{Conclusion}

This paper presents a novel approach to dissect and quantify enthusiasm in computer science (CS) education, shedding light on the nuanced interplay between age, gender, and interest in the field. By developing a novel questionnaire grounded in the Person-Object Theory of Interest (POI), this study not only addresses a critical gap in the literature but also introduces a tool for educators and researchers to assess the impact of pedagogical strategies on students' CS enthusiasm.

The contributions of this work include: a novel measurement instrument, specifically designed to capture the potential of CS interventions to raise enthusiasm, as well as empirical evidence that delineates how age significantly influences students' interest in CS, identifying crucial developmental breakpoints with profound implications for curriculum design and instructional approaches.

Despite this study's contributions, we acknowledge its limitations, including the demographic and geographic scope of the participant pool. Future research should therefore aim to:
\begin{itemize}
    \item Conduct longitudinal analyses to track the evolution of students' CS interest over time, offering insights into the long-term efficiency of educational interventions in raising enthusiasm for CS. A discussion on the implications of this research for lifelong learning and career development in CS could be intriguing. It would be beneficial to explore how early interest, or lack thereof, impacts students' long-term engagement with CS, their career choices, and their continuous skill development in an ever-evolving field.
    \item Explore the impact of diverse teaching methodologies on both gender and age groups to refine and personalize CS education strategies. For instance, interdisciplinary and project-based learning, gamification, and the integration of CS across different subjects may have varying effects on student engagement. Reflecting on how these approaches align with students' developmental stages and evolving interests could yield valuable insights for educators.
    \item Expand the research to include a broader range of educational contexts, previous major interests, and cultural backgrounds, enhancing the generalizability of the findings. A discussion of the relationship between CS enthusiasm and interest in other STEM and non-STEM fields could reveal patterns or shared challenges. Given the interdisciplinary nature of many modern scientific and engineering challenges, exploring how enthusiasm for CS correlates with or diverges from trends in other subjects could offer insights into how to foster a more integrated education approach.
\end{itemize}

In summary, this study not only advances our understanding of the factors influencing interest in CS among young learners but also equips stakeholders with a tool to measure and enhance student enthusiasm. As we look to the future, the insights garnered here lay a solid foundation for fostering a more diverse, engaging, and effective CS education landscape, ultimately contributing to the diversification and growth of the global CS workforce.

\bibliography{references}

@article{kimberlin2008validity,
    author = {Kimberlin, Carole L. and Winterstein, Almut G.},
    title = {Validity and reliability of measurement instruments used in research},
    journal = {American Journal of Health-System Pharmacy},
    volume = {65},
    number = {23},
    pages = {2276-2284},
    year = {2008},
    doi = {10.2146/ajhp070364},
    eprint = {https://academic.oup.com/ajhp/article-pdf/65/23/2276/28304597/ajhp2276.pdf},
}

@article{scholters2011what,
title = {What makes a measurement instrument valid and reliable?},
journal = {Injury},
volume = {42},
number = {3},
pages = {236-240},
year = {2011},
note = {Assessing Patient Outcomes},
issn = {0020-1383},
doi = {https://doi.org/10.1016/j.injury.2010.11.042},
author = {Vanessa A. Scholtes and Caroline B. Terwee and Rudolf W. Poolman}
}

@article{bollen2002latent,
   author = "Bollen, Kenneth A.",
   title = "Latent Variables in Psychology and the Social Sciences", 
   journal= "Annual Review of Psychology",
   year = "2002",
   volume = "53",
   number = "Volume 53, 2002",
   pages = "605-634",
   doi = "https://doi.org/10.1146/annurev.psych.53.100901.135239",
   url = "https://www.annualreviews.org/content/journals/10.1146/annurev.psych.53.100901.135239",
   publisher = "Annual Reviews"
  }

@article{schermelleh2003evaluating,
  title={Evaluating the Fit of Structural Equation Models: Tests of Significance and Descriptive Goodness-of-Fit Measures},
  author={Schermelleh-Engel, Karin and Moosbrugger, Helfried and M{\"u}ller, Hans and others},
  journal={Methods of Psychological Research},
  volume={8},
  number={2},
  pages={23--74},
  year={2003}
}

@article{rigdon1996cfi,
author = {Edward E. Rigdon},
title = {CFI versus RMSEA: A comparison of two fit indexes for structural equation modeling},
journal = {Structural Equation Modeling: A Multidisciplinary Journal},
volume = {3},
number = {4},
pages = {369--379},
year = {1996},
publisher = {Routledge},
doi = {10.1080/10705519609540052},
}

@article{bentler1990comparative,
  title={Comparative fit indexes in structural models.},
  author={Bentler, Peter M},
  journal={Psychological bulletin},
  volume={107},
  number={2},
  pages={238--246},
  year={1990},
  publisher={American Psychological Association}
}

@article{master2025causes,
  title={Causes and consequences of stereotypes: interest stereotypes reduce adolescent girls’ motivation to enroll in computer science classes},
  author={Master, Allison and Alexander, Taylor and Thompson, Jennifer and Fan, Weihua and Meltzoff, Andrew N and Cheryan, Sapna},
  journal={Journal of Research on Technology in Education},
  volume={57},
  number={1},
  pages={56--83},
  year={2025},
  publisher={Taylor \& Francis}
}

@article{tavakol2011making,
  title={Making sense of Cronbach's alpha},
  author={Tavakol, Mohsen and Dennick, Reg},
  journal={International journal of medical education},
  volume={2},
  pages={53--55},
  year={2011},
  publisher={IJME},
  doi={https://doi.org/10.5116/ijme.4dfb.8dfd}
}

@article{hidi2006four,
  title={The four-phase model of interest development},
  author={Hidi, Suzanne and Renninger, K Ann},
  journal={Educational psychologist},
  volume={41},
  number={2},
  pages={111--127},
  year={2006},
  publisher={Taylor \& Francis}
}

@article{krapp2007educational,
  title={An educational--psychological conceptualisation of interest},
  author={Krapp, Andreas},
  journal={International journal for educational and vocational guidance},
  volume={7},
  number={1},
  pages={5--21},
  year={2007},
  publisher={Springer}
}

@inproceedings{schiefele1983principles,
  title={Principles of an educational theory of interest},
  author={Schiefele, Hans and Krapp, Andreas and Prenzel, Manfred and Heiland, Andreas and Kasten, Hartmut},
  booktitle={7th Meeting of the International Society for the Study of Behavioral Development in Munich},
  year={1983}
}

@article{happe2021effective,
  title={Effective measures to foster girls’ interest in secondary computer science education},
  author={Happe, Lucia and Buhnova, Barbora and Koziolek, Anne and Wagner, Ingo},
  journal={Education and Information Technologies},
  volume={26},
  number={3},
  pages={2811--2829},
  year={2021},
  publisher={Springer}
}

@inproceedings{vainionpaeae2019girls,
author = {Vainionpää, Fanny and Kinnula, Marianne and Iivari, Netta and Molin-Juustila, Tonja},
title = {GIRLS' CHOICE - WHY WON'T THEY PICK IT?},
year = {2019},
isbn = {9781733632508},
url = {https://aisel.aisnet.org/ecis2019\_rp/31},
booktitle = {Proceedings of the 27th European Conference on Information Systems (ECIS)},
location = {Stockholm and Uppsala, Sweden},
series = {ECIS '19}
}

@inproceedings{vela2018matters,
  title={What matters to my future: STEM int-her-est and expectations},
  author={Vela, Katherine N and Bicer, Ali and Capraro, Robert M and Barroso, Luciana R and Caldwell, Cassidy},
  booktitle={2018 IEEE Frontiers in Education Conference (FIE)},
  pages={1--7},
  year={2018},
  organization={IEEE}
}

@article{rotgans2017interest,
  title={Interest development: Arousing situational interest affects the growth trajectory of individual interest},
  author={Rotgans, Jerome I and Schmidt, Henk G},
  journal={Contemporary Educational Psychology},
  volume={49},
  pages={175--184},
  year={2017},
  publisher={Elsevier}
}

@article{dubow2016male,
  title={Male allies: Motivations and barriers for participating in diversity initiatives in the technology workplace},
  author={DuBow, Wendy M and Ashcraft, Catherine},
  journal={International Journal of Gender, Science and Technology},
  volume={8},
  number={2},
  pages={160--180},
  year={2016}
}

@inproceedings{marquardt2024crossing,
author = {Marquardt, Kai and Happe, Lucia},
title = {Crossing Borders: The Beauty of Computer Science Beyond Disciplinary Borders},
year = {2024},
publisher = {Association for Computing Machinery},
address = {New York, NY, USA},
location = {Madrid, Spain},
series = {womENcourage™ 2024},
url={https://womencourage.acm.org/2024/wp-content/uploads/2024/06/womencourage2024-posters-final5.pdf},
 urldate = {22/07/2024},
}

@article{happe2025authentic,
title = {Authentic Interdisciplinary Online Courses for More Diversity in Computer Science},
journal = {Journal of Systems and Software},
volume = {219},
year = {2025},
pages = {112240},
issn = {0164-1212},
author = {Happe, Lucia and \textbf{Kai Marquardt}},
doi={https://doi.org/10.1016/j.jss.2024.112240}
}

@techreport{royalsociety2025,
  author = {{The Royal Society}},
  title = {System upgrade required: Creating opportunities in computing education},
  institution = {The Royal Society},
  year = {2025},
  month = {September},
  url = {https://royalsociety.org/news-resources/projects/computing-education/},
  urldate={2026-02-14},
  isbn = {978-1-78252-802-9}
}

@misc{spss2025,
title={IBM SPSS Statistics for Windows (Version 29.0) [Computer software]},
 author={{IBM Corp.}},
 year={2025}
}

@article{kekelis2005hurdles,
 ISSN = {01609009, 15360334},
 note = {http://www.jstor.org/stable/4137438},
 author = {Linda S. Kekelis and Rebecca Wepsic Ancheta and Etta Heber},
 journal = {Frontiers: A Journal of Women Studies},
 number = {1},
 pages = {99--109},
 publisher = {University of Nebraska Press},
 title = {Hurdles in the Pipeline: Girls and Technology Careers},
 urldate = {2024-07-22},
 volume = {26},
 year = {2005}
}

@article{eccles1994understanding,
author = {Jacquelynne S. Eccles},
title ={Understanding Women's Educational And Occupational Choices: Applying the Eccles et al. Model of Achievement-Related Choices},
journal = {Psychology of Women Quarterly},
volume = {18},
number = {4},
pages = {585-609},
year = {1994},
doi = {10.1111/j.1471-6402.1994.tb01049.x}
}

@article{dasgupta2014girls,
  author = {Nilanjana Dasgupta and Jane G. Stout},
title ={Girls and Women in Science, Technology, Engineering, and Mathematics: STEMing the Tide and Broadening Participation in STEM Careers},
journal = {Policy Insights from the Behavioral and Brain Sciences},
volume = {1},
number = {1},
pages = {21-29},
year = {2014},
doi = {10.1177/2372732214549471}
}

@article{super1980life,
  author={Super, Donald E},
title = {A life-span, life-space approach to career development},
journal = {Journal of Vocational Behavior},
volume = {16},
number = {3},
pages = {282-298},
year = {1980},
issn = {0001-8791},
doi = {https://doi.org/10.1016/0001-8791(80)90056-1}
}

@misc{jamovi2025,
 title={jamovi (Version 2.6) [Computer Software]},
 author={{The jamovi project}},
 year={2025},
 url ={https://www.jamovi.org}
}

@article{ruthotto2020lurking,
  title={Lurking and participation in the virtual classroom: The effects of gender, race, and age among graduate students in computer science},
  author={Ruthotto, Isabel and Kreth, Quintin and Stevens, Jillian and Trively, Clare and Melkers, Julia},
  journal={Computers \& Education},
  volume={151},
  pages={103854},
  year={2020},
  publisher={Elsevier}
}

@article{palma2001women,
  title={Why women avoid computer science},
  author={Palma, Paul De},
  journal={Communications of the ACM},
  volume={44},
  number={6},
  pages={27--29},
  year={2001},
  publisher={ACM New York, NY, USA}
}

@inproceedings{gursch2022coding,
  title={Coding initiative provides different approaches to inspire girls for programming},
  author={Gursch, Sarina},
  booktitle={International Conference on Computers in Education (ICCE)},
  year={2022},
  url={https://library.apsce.net/index.php/ICCE/article/view/4551}
}

@misc{r2023,
title={R: A language and environment for statistical computing [Computer software]},
year={2023},
author={{R Core Team}}
}

@article{gorbacheva2019directions,
  title={Directions for research on gender imbalance in the IT profession},
  author={Gorbacheva, Elena and Beekhuyzen, Jenine and vom Brocke, Jan and Becker, J{\"o}rg},
  journal={European Journal of Information Systems},
  volume={28},
  number={1},
  pages={43--67},
  year={2019},
  publisher={Taylor \& Francis},
  doi={https://doi.org/10.1080/0960085X.2018.1495893}
}

@article{kirikkaya2011grade,
  title={Grade 4 to 8 primary school students' attitudes towards science: Science enthusiasm},
  author={Kirikkaya, Esma Bulu},
  journal={Educational Research and Reviews},
  volume={6},
  number={4},
  pages={374--382},
  year={2011},
  publisher={Academic Journals}
}

@book{csikszentmihalyi1990flow,
  title={Flow: The psychology of optimal experience},
  author={Csikszentmihalyi, Mihaly and Csikzentmihaly, Mihaly},
  volume={1990},
  year={1990},
  publisher={Harper \& Row New York}
}

@inproceedings{singh2020towards,
  title={Towards enthusiasm prediction of Portuguese school's students towards higher education in realtime},
  author={Singh, Mandeep and Verma, Chaman and Kumar, Rajiv and Juneja, Pamela},
  booktitle={2020 International Conference on Computation, Automation and Knowledge Management (ICCAKM)},
  pages={421--425},
  year={2020},
  organization={IEEE}
}

@article{alpay2008student,
  title={Student enthusiasm for engineering: charting changes in student aspirations and motivation},
  author={Alpay, E and Ahearn, AL and Graham, RH and Bull, AMJ},
  journal={European Journal of Engineering Education},
  volume={33},
  number={5-6},
  pages={573--585},
  year={2008},
  publisher={Taylor \& Francis}
}

@article{ng2020engaging,
  title={Engaging high school girls in interdisciplinary STEAM},
  author={Ng, Wan and Fergusson, Jennifer},
  journal={Science Education International},
  volume={31},
  number={3},
  pages={283--294},
  year={2020}
}

@article{chipman2018evaluating,
  title={Evaluating computer science camp topics in increasing girls' confidence in computer science},
  author={Chipman, Hannah and Adams, Haley and Sanders, Betsy Williams and Larkins, D Brian},
  journal={Journal of Computing Sciences in Colleges},
  volume={33},
  number={5},
  pages={70--78},
  year={2018},
  publisher={Consortium for Computing Sciences in Colleges}
}

@incollection{friend2017girls,
  title={Girls' Interest in Computing: Types and Persistence},
  author={Friend, Michelle},
  year={2017},
  publisher={Philadelphia, PA: International Society of the Learning Sciences.}
}

@phdthesis{beumann2017versuch,
  title={Versuch{\'{}} s doch mal},
  author={Beumann, Sarah},
  school  = "Ruhr-Universit{\"a}t Bochum",
  address = "Bochum, Germany",
  month   = jan,
  year={2017}
}

@phdthesis{zehren2009forschendes,
  title={Forschendes Experimentieren im Sch{\"u}lerlabor},
  author={Zehren, Walter},
  school  = "Universit{\"a}t des Saarlandes",
  address = "Saarbr{\"u}cken, Germany",
  month   = jul,
  year={2009}
}

@article{harackiewicz2016interest,
  title={Interest matters: The importance of promoting interest in education},
  author={Harackiewicz, Judith M and Smith, Jessi L and Priniski, Stacy J},
  journal={Policy insights from the behavioral and brain sciences},
  volume={3},
  number={2},
  pages={220--227},
  year={2016},
  publisher={SAGE Publications Sage CA: Los Angeles, CA}
}

@article{palmer2017using,
  title={Using situational interest to enhance individual interest and science-related behaviours},
  author={Palmer, David and Dixon, Jeanette and Archer, Jennifer},
  journal={Research in Science Education},
  volume={47},
  number={4},
  pages={731--753},
  year={2017},
  publisher={Springer}
}

@inproceedings{ericson2012effective,
  title={Effective and sustainable computing summer camps},
  author={Ericson, Barbara and McKlin, Tom},
  booktitle={Proceedings of the 43rd ACM technical symposium on Computer Science Education},
  pages={289--294},
  year={2012}
}

@inproceedings{katterfeldt2019effects,
  title={Effects of Physical Computing Workshops on Girls' Attitudes towards Computer Science},
  author={Katterfeldt, Eva-Sophie and Dittert, Nadine and Ghose, Sobin and Bernin, Arne and Daeglau, Mareike},
  booktitle={Proceedings of the FabLearn Europe 2019 Conference},
  pages={1--3},
  year={2019}
}

@inproceedings{jenson2017gender,
  title={Gender and Game Making: Attitudes, Competencies and Computational Thinking},
  author={Jenson, Jennifer and Black, Karen},
 booktitle={Proceedings of the DIGRA 2017 Conference},
  year={2017}
}

@article{muller2007skalen,
  title={Skalen zur motivationalen Regulation beim Lernen von Sch{\"u}lerinnen und Sch{\"u}lern},
  author={M{\"u}ller, Florian H and Hanfstingl, Barbara and Andreitz, Irina},
  journal={Adaptierte und erg{\"a}nzte Version des Academic},
  volume={242},
  year={2007}
}

@incollection{haussler2007lasst,
  title={Wie l{\"a}sst sich der Lernerfolg messen?},
  author={H{\"a}ussler, Peter},
  booktitle={Physikdidaktik},
  pages={249--294},
  year={2007},
  publisher={Springer}
}

@book{engeln2004schulerlabors,
  title={Sch{\"u}lerlabors: authentische, aktivierende Lernumgebungen als M{\"o}glichkeit, Interesse an Naturwissenschaften und Technik zu wecken},
  author={Engeln, Katrin},
  year={2004},
  publisher={Logos-Verlag}
}

@misc{Table,
 title={The Literature Survey 2016-2019},
  howpublished = {https://docs.google.com/spreadsheets/d/1UtMRgpqpfoaJjOnv6hOYfLYMpGp8gLdb7PD1qEt4HCM/edit?usp=sharing}
}

@article{master2017,
  title={Programming experience promotes higher STEM motivation among first-grade girls},
  author={Master, Allison and Cheryan, Sapna and Moscatelli, Adriana and Meltzoff, Andrew N},
  journal={Journal of experimental child psychology},
  volume={160},
  pages={92--106},
  year={2017},
  publisher={Elsevier}
}

@article{maccallum1999sample,
  title={Sample size in factor analysis.},
  author={MacCallum, Robert C and Widaman, Keith F and Zhang, Shaobo and Hong, Sehee},
  journal={Psychological methods},
  volume={4},
  number={1},
  pages={84},
  year={1999},
  publisher={American Psychological Association}
}

@inproceedings{henry2018,
  title={Perceptions of computer science among children after a hands-on activity: A pilot study},
  author={Henry, Julie and Dumas, Bruno},
  booktitle={2018 IEEE Global Engineering Education Conference (EDUCON)},
  pages={1811--1817},
  year={2018},
  organization={IEEE}
}

@inproceedings{burns2017,
  title={Empathy in middle school engineering design process},
  author={Burns, Henriette D and Lesseig, Kristin},
  booktitle={2017 IEEE Frontiers in Education Conference (FIE)},
  pages={1--4},
  year={2017},
  organization={IEEE}
}

@article{sabin2017,
  title={Summer learning experience for girls in grades 7--9 boosts confidence and interest in computing careers},
  author={Sabin, Mihaela C and Deloge, Rosabel and Smith, Adrienne and DuBow, Wendy},
  journal={Journal of Computing Sciences in Colleges},
  year={2017},
  publisher={ACM (Association for Computing Machinery)}
}

@article{tellhed2022sure,
  title={Sure I can code (but do I want to?). Why boys' and girls’ programming beliefs differ and the effects of mandatory programming education},
  author={Tellhed, Una and Bj{\"o}rklund, Fredrik and Strand, Kalle Kallio},
  journal={Computers in Human Behavior},
  volume={135},
  pages={107370},
  year={2022},
  publisher={Elsevier}
}

@article{rotgans2015validation,
  title={Validation study of a general subject-matter interest measure: The Individual Interest Questionnaire (IIQ)},
  author={Rotgans, Jerome I},
  journal={Health Professions Education},
  volume={1},
  number={1},
  pages={67--75},
  year={2015},
  publisher={Elsevier}
}

@book{DeciR85,
  author = {Deci, Edward L. and Ryan, Richard M.},
  ee = {https://www.wikidata.org/entity/Q54187435},
  isbn = {978-1-4899-2271-7},
  pages = {1-371},
  publisher = {Springer},
  series = {Perspectives in Social Psychology},
  title = {Intrinsic Motivation and Self-Determination in Human Behavior},
  year = 1985
}

@article{renninger2011revisiting,
  title={Revisiting the conceptualization, measurement, and generation of interest},
  author={Renninger, K Ann and Hidi, Suzanne},
  journal={Educational psychologist},
  volume={46},
  number={3},
  pages={168--184},
  year={2011},
  publisher={Taylor \& Francis}
}

@article{krapp2002structural,
  title={Structural and dynamic aspects of interest development: Theoretical considerations from an ontogenetic perspective},
  author={Krapp, Andreas},
  journal={Learning and instruction},
  volume={12},
  number={4},
  pages={383--409},
  year={2002},
  publisher={Elsevier}
}

@article{knogler2015situational,
  title={How situational is situational interest? Investigating the longitudinal structure of situational interest},
  author={Knogler, Maximilian and Harackiewicz, Judith M and Gegenfurtner, Andreas and Lewalter, Doris},
  journal={Contemporary Educational Psychology},
  volume={43},
  pages={39--50},
  year={2015},
  publisher={Elsevier}
}

@report{grosch2020mint,
  title={MINT-Interesse bei Kindern steigern: Ein Feldexperiment an Volksschulen in {\"O}sterreich},
  author={Grosch, Kerstin and H{\"a}ckl, Simone and Kocher, Martin G and Bauer, Christian},
  year={2020}
}

@book{wommel2016enthusiasmus,
  title={Was ist Enthusiasmus?},
  author={W{\"o}mmel, Kristin and W{\"o}mmel, Kristin},
  year={2016},
  publisher={Springer}
}

@inproceedings{gopalan2017review,
  title={A review of the motivation theories in learning},
  author={Gopalan, Valarmathie and Bakar, Juliana Aida Abu and Zulkifli, Abdul Nasir and Alwi, Asmidah and Mat, Ruzinoor Che},
  booktitle={Aip conference proceedings},
  volume={1891},
  number={1},
  year={2017},
  organization={AIP Publishing}
}

@inproceedings{aivaloglou2019early,
  title={Early programming education and career orientation: the effects of gender, self-efficacy, motivation and stereotypes},
  author={Aivaloglou, Efthimia and Hermans, Felienne},
  booktitle={Proceedings of the 50th ACM technical symposium on computer science education},
  pages={679--685},
  year={2019}
}

@article{colley2003age,
  title={Age and gender differences in computer use and attitudes among secondary school students: what has changed?},
  author={Colley, Ann and Comber, Chris},
  journal={Educational research},
  volume={45},
  number={2},
  pages={155--165},
  year={2003},
  publisher={Taylor \& Francis}
}

@article{holm1979simple,
  title={A simple sequentially rejective multiple test procedure},
  author={Holm, Sture},
  journal={Scandinavian journal of statistics},
  pages={65--70},
  year={1979},
  publisher={JSTOR}
}

@article{richardson2011eta,
  title={Eta squared and partial eta squared as measures of effect size in educational research},
  author={Richardson, John TE},
  journal={Educational research review},
  volume={6},
  number={2},
  pages={135--147},
  year={2011},
  publisher={Elsevier}
}

@book{cohen1988statistical,
  title={Statistical power analysis for the behavioral sciences},
  author={Cohen, Jacob},
  year={1988},
  publisher={Routledge}
}

@article{fabrigar1999evaluating,
  title={Evaluating the use of exploratory factor analysis in psychological research.},
  author={Fabrigar, Leandre R and Wegener, Duane T and MacCallum, Robert C and Strahan, Erin J},
  journal={Psychological methods},
  volume={4},
  number={3},
  pages={272},
  year={1999},
  publisher={American Psychological Association}
}

@article{kim2014analysis,
  title={Analysis of variance (ANOVA) comparing means of more than two groups},
  author={Kim, Hae-Young},
  journal={Restorative dentistry \& endodontics},
  volume={39},
  number={1},
  pages={74},
  year={2014},
  publisher={Korean Academy of Conservative Dentistry}
}

@article{glynn2011science,
  title={Science motivation questionnaire II: Validation with science majors and nonscience majors},
  author={Glynn, Shawn M and Brickman, Peggy and Armstrong, Norris and Taasoobshirazi, Gita},
  journal={Journal of research in science teaching},
  volume={48},
  number={10},
  pages={1159--1176},
  year={2011},
  publisher={Wiley Online Library}
}

@article{krapp1999interest,
  title={Interest, motivation and learning: An educational-psychological perspective},
  author={Krapp, Andreas},
  journal={European journal of psychology of education},
  volume={14},
  pages={23--40},
  year={1999},
  publisher={Springer}
}

@article{schiefele1991interest,
  title={Interest, learning, and motivation},
  author={Schiefele, Ulrich},
  journal={Educational psychologist},
  volume={26},
  number={3-4},
  pages={299--323},
  year={1991},
  publisher={Taylor \& Francis}
}

@phdthesis{pawek2009schulerlabore,
  title={Sch{\"u}lerlabore als interessef{\"o}rdernde au{\ss}erschulische Lernumgebungen f{\"u}r Sch{\"u}lerinnen und Sch{\"u}ler aus der Mittel-und Oberstufe},
  author={Pawek, Christoph},
  year={2009}
}

@article{kunter2011teacher,
  title={Teacher enthusiasm: Dimensionality and context specificity},
  author={Kunter, Mareike and Frenzel, Anne and Nagy, Gabriel and Baumert, J{\"u}rgen and Pekrun, Reinhard},
  journal={Contemporary Educational Psychology},
  volume={36},
  number={4},
  pages={289--301},
  year={2011},
  publisher={Elsevier}
}

@article{cheryan2015cultural,
  title={Cultural stereotypes as gatekeepers: Increasing girls’ interest in computer science and engineering by diversifying stereotypes},
  author={Cheryan, Sapna and Master, Allison and Meltzoff, Andrew N},
  journal={Frontiers in psychology},
  pages={49},
  year={2015},
  publisher={Frontiers}
}

@inproceedings{lewis2016don,
  title={" I Don't Code All Day" Fitting in Computer Science When the Stereotypes Don't Fit},
  author={Lewis, Colleen M and Anderson, Ruth E and Yasuhara, Ken},
  booktitle={Proceedings of the 2016 ACM conference on international computing education research},
  pages={23--32},
  year={2016}
}

@article{norman2010likert,
  title={Likert scales, levels of measurement and the “laws” of statistics},
  author={Norman, Geoff},
  journal={Advances in health sciences education},
  volume={15},
  pages={625--632},
  year={2010},
  publisher={Springer},
doi={https://doi.org/10.1007/s10459-010-9222-y}
}

@article{joshi2015likert,
  title={Likert scale: Explored and explained},
  author={Joshi, Ankur and Kale, Saket and Chandel, Satish and Pal, D Kumar},
  journal={British journal of applied science \& technology},
  volume={7},
  number={4},
  pages={396--403},
  year={2015},
doi={https://doi.org/10.9734/BJAST/2015/14975}
}

@article{mensah2019computer,
  title={Computer self-efficacy and e-government service adoption: the moderating role of age as a demographic factor},
  author={Mensah, Isaac Kofi and Mi, Jianing},
  journal={International Journal of Public Administration},
  volume={42},
  number={2},
  pages={158--167},
  year={2019},
  publisher={Taylor \& Francis}
}

@article{chu2010family,
  title={How family support and Internet self-efficacy influence the effects of e-learning among higher aged adults--Analyses of gender and age differences},
  author={Chu, Regina Ju-chun},
  journal={Computers \& Education},
  volume={55},
  number={1},
  pages={255--264},
  year={2010},
  publisher={Elsevier}
}

@article{master2021gender,
  title={Gender stereotypes about interests start early and cause gender disparities in computer science and engineering},
  author={Master, Allison and Meltzoff, Andrew N and Cheryan, Sapna},
  journal={Proceedings of the National Academy of Sciences},
  volume={118},
  number={48},
  pages={e2100030118},
  year={2021},
  publisher={National Acad Sciences}
}

@article{schorr2019pipped,
  title={Pipped at the post: knowledge gaps and expected low parental IT competence ratings affect young women’s awakening interest in professional careers in information science},
  author={Schorr, Angela},
  journal={Frontiers in psychology},
  volume={10},
  pages={968},
  year={2019},
  publisher={Frontiers}
}

@article{blankenburg2015naturwissenschaftliche,
  title={Naturwissenschaftliche Wettbewerbe--Was kann junge Sch{\"u}lerinnen und Sch{\"u}ler zur Teilnahme motivieren?},
  author={Blankenburg, Janet Susan and H{\"o}ffler, Tim Niclas and Parchmann, Ilka},
  journal={Zeitschrift f{\"u}r Didaktik der Naturwissenschaften},
  volume={21},
  number={1},
  pages={141--153},
  year={2015},
  publisher={Springer}
}

@article{ertl2014stereotye,
  title={Stereotype als Einflussfaktoren auf die Motivation und die Einsch{\"a}tzung der eigenen F{\"a}higkeiten bei Studentinnen in MINT-F{\"a}chern},
  author={Ertl, Bernhard and Luttenberger, Silke and Paechter, Manuela},
  journal={Gruppendynamik und Organisationsberatung},
  volume={45},
  number={4},
  pages={419--440},
  year={2014},
  publisher={Springer}
}

@phdthesis{glowinski2007schuelerlabore,
  title={Sch{\"u}lerlabore im Themenbereich Molekularbiologie als Interesse f{\"o}rdernde Lernumgebungen},
  author={Glowinski, Ingrid},
  year={2007}
}

@article{haselmeier2019interesse,
  title={Interesse an Informatik und Informatikselbstkonzept zu Beginn der Sekundarstufe I des Gymnasiums},
  author={Haselmeier, Kathrin and Humbert, Ludger and Killich, Klaus and M{\"u}ller, Dorothee},
  journal={Informatik f{\"u}r alle},
  year={2019},
  publisher={Gesellschaft f{\"u}r Informatik}
}

@article{kukul2017computer,
  title={Computer programming self-efficacy scale (CPSES) for secondary school students: Development, validation and reliability},
  author={Kukul, Volkan and G{\"o}k{\c{c}}earslan, {\c{S}}ahin and G{\"u}nbatar, Mustafa Serkan},
  journal={Egitim Teknolojisi Kuram ve Uygulama},
  year={2017},
  volume={7},
  number={1},
  pages={158--179}
}

@article{outlay2017getting,
  title={Getting IT together: A longitudinal look at linking girls' interest in IT careers to lessons taught in middle school camps},
  author={Outlay, Christina N and Platt, Alana J and Conroy, Kacie},
  journal={ACM Transactions on Computing Education (TOCE)},
  volume={17},
  number={4},
  pages={1--17},
  year={2017},
  publisher={ACM New York, NY, USA}
}

@inproceedings{theodoropoulos2018computing,
  title={Computing in the physical world engages students: impact on their attitudes and self-efficacy towards computer science through robotic activities},
  author={Theodoropoulos, Anastasios and Leon, Prokopis and Antoniou, Angeliki and Lepouras, George},
  booktitle={Proceedings of the 13th Workshop in Primary and Secondary Computing Education},
  pages={1--4},
  year={2018}
}

@article{dunn2014alpha,
  title={From alpha to omega: A practical solution to the pervasive problem of internal consistency estimation},
  author={Dunn, Thomas J and Baguley, Thom and Brunsden, Vivienne},
  journal={British journal of psychology},
  volume={105},
  number={3},
  pages={399--412},
  year={2014},
  publisher={Wiley Online Library}
}

@article{pintrich1990motivational,
  title={Motivational and self-regulated learning components of classroom academic performance.},
  author={Pintrich, Paul R and De Groot, Elisabeth V},
  journal={Journal of educational psychology},
  volume={82},
  number={1},
  pages={33--40},
  year={1990},
  publisher={American Psychological Association}
}

@inproceedings{kaloti2015students,
  title={Students' attitudes and motivation during robotics activities},
  author={Kaloti-Hallak, Fatima and Armoni, Michal and Ben-Ari, Mordechai},
  booktitle={Proceedings of the Workshop in Primary and Secondary Computing Education},
  pages={102--110},
  year={2015}
}

@article{korkmaz2017validity,
  title={A validity and reliability study of the computational thinking scales (CTS)},
  author={Korkmaz, {\"O}zgen and {\c{C}}akir, Recep and {\"O}zden, M Ya{\c{s}}ar},
  journal={Computers in human behavior},
  volume={72},
  pages={558--569},
  year={2017},
  publisher={Elsevier}
}

@article{unfried2015development,
  title={The development and validation of a measure of student attitudes toward science, technology, engineering, and math (S-STEM)},
  author={Unfried, Alana and Faber, Malinda and Stanhope, Daniel S and Wiebe, Eric},
  journal={Journal of Psychoeducational Assessment},
  volume={33},
  number={7},
  pages={622--639},
  year={2015},
  publisher={SAGE Publications Sage CA: Los Angeles, CA}
}

@inproceedings{faber2013students,
  title={Student attitudes toward STEM: The development of upper elementary school and middle/high school student surveys},
  author={Faber, Malinda and Unfried, Alana and Wiebe, Eric N and Corn, Jeni and Townsend, LW and Collins, Tracey L},
  booktitle={the Proceedings of the 120th American Society of Engineering Education Conference},
  year={2013}
}

@article{erkut2005schools,
  title={4 Schools for WIE. Evaluation Report.},
  author={Erkut, Sumru and Marx, Fern},
  journal={Wellesley Centers for Women},
  year={2005},
  publisher={ERIC}
}

@article{holmes2018integrated,
  title={An integrated analysis of school students’ aspirations for STEM careers: Which student and school factors are most predictive?},
  author={Holmes, Kathryn and Gore, Jennifer and Smith, Max and Lloyd, Adam},
  journal={International Journal of Science and Mathematics Education},
  volume={16},
  number={4},
  pages={655--675},
  year={2018},
  publisher={Springer}
}

@article{glass1972consequences,
  title={Consequences of failure to meet assumptions underlying the fixed effects analyses of variance and covariance},
  author={Glass, Gene V and Peckham, Percy D and Sanders, James R},
  journal={Review of educational research},
  volume={42},
  number={3},
  pages={237--288},
  year={1972},
  publisher={Sage Publications Sage CA: Thousand Oaks, CA}
}

@article{harwell1992summarizing,
  title={Summarizing Monte Carlo results in methodological research: The one-and two-factor fixed effects ANOVA cases},
  author={Harwell, Michael R and Rubinstein, Elaine N and Hayes, William S and Olds, Corley C},
  journal={Journal of educational statistics},
  volume={17},
  number={4},
  pages={315--339},
  year={1992},
  publisher={Sage Publications Sage CA: Los Angeles, CA}
}

@article{comber1997effects,
  title={The effects of age, gender and computer experience upon computer attitudes},
  author={Comber, Chris and Colley, Ann and Hargreaves, David J and Dorn, Lisa},
  journal={Educational research},
  volume={39},
  number={2},
  pages={123--133},
  year={1997},
  publisher={Taylor \& Francis}
}

@article{master2016computing,
  title={Computing whether she belongs: Stereotypes undermine girls’ interest and sense of belonging in computer science.},
  author={Master, Allison and Cheryan, Sapna and Meltzoff, Andrew N},
  journal={Journal of educational psychology},
  volume={108},
  number={3},
  pages={424},
  year={2016},
  publisher={American Psychological Association}
}

@article{french2018females,
author = {French, Jean H. and Crouse, Hailey},
title = {Using early intervention to increase female interest in computing sciences},
year = {2018},
issue_date = {December 2018},
publisher = {Consortium for Computing Sciences in Colleges},
address = {Evansville, IN, USA},
volume = {34},
number = {2},
issn = {1937-4771},
journal = {J. Comput. Sci. Coll.},
month = {dec},
pages = {133–140},
numpages = {8}
}

@article{fandakova2021states,
  title={States of curiosity and interest enhance memory differently in adolescents and in children},
  author={Fandakova, Yana and Gruber, Matthias J},
  journal={Developmental Science},
  volume={24},
  number={1},
  pages={e13005},
  year={2021},
  publisher={Wiley Online Library}
}

@article{elhamamsy2023primary,
  title={How are primary school computer science curricular reforms contributing to equity? Impact on student learning, perception of the discipline, and gender gaps},
  author={El-Hamamsy, Laila and Bruno, Barbara and Audrin, Catherine and Chevalier, Morgane and Avry, Sunny and Zufferey, Jessica Dehler and Mondada, Francesco},
  journal={International Journal of STEM Education},
  volume={10},
  number={1},
  pages={60},
  year={2023},
  publisher={Springer}
}

@inproceedings{mcclure2017stem,
  title={STEM starts early: Grounding science, technology, engineering, and math education in early childhood.},
  author={McClure, Elisabeth R and Guernsey, Lisa and Clements, Douglas H and Bales, Susan Nall and Nichols, Jennifer and Kendall-Taylor, Nat and Levine, Michael H},
  booktitle={Joan Ganz Cooney center at sesame workshop},
  year={2017},
  organization={ERIC}
}

@article{cavanagh2007puberty,
author = {Shannon E. Cavanagh and Catherine Riegle-Crumb and Robert Crosnoe},
title ={Puberty and the Education of Girls},
journal = {Social Psychology Quarterly},
volume = {70},
number = {2},
pages = {186-198},
year = {2007},
doi = {10.1177/019027250707000207},
note ={PMID: 20216926},
URL = {https://doi.org/10.1177/019027250707000207},
eprint = {https://doi.org/10.1177/019027250707000207}
}

@article{marceau2011individual,
  title={Individual differences in boys' and girls' timing and tempo of puberty: modeling development with nonlinear growth models.},
  author={Marceau, Kristine and Ram, Nilam and Houts, Renate M and Grimm, Kevin J and Susman, Elizabeth J},
  journal={Developmental psychology},
  volume={47},
  number={5},
  pages={1389},
  year={2011},
  publisher={American Psychological Association}
}

@article{beyer2014women,
  title={Why are women underrepresented in Computer Science? Gender differences in stereotypes, self-efficacy, values, and interests and predictors of future CS course-taking and grades},
  author={Beyer, Sylvia},
  journal={Computer Science Education},
  volume={24},
  number={2-3},
  pages={153--192},
  year={2014},
  publisher={Taylor \& Francis}
}

@inproceedings{berg2023elaborating,
author = {Berg, Pamina Maria and Knobelsdorf, Maria},
title = {Elaborating Student Engagement in Compulsory K-12 Computer Science Class},
year = {2023},
isbn = {9798400708510},
publisher = {Association for Computing Machinery},
address = {New York, NY, USA},
url = {https://doi.org/10.1145/3605468.3609762},
doi = {10.1145/3605468.3609762},
booktitle = {Proceedings of the 18th WiPSCE Conference on Primary and Secondary Computing Education Research},
articleno = {16},
numpages = {4},
location = {Cambridge, United Kingdom},
series = {WiPSCE '23}
}

\end{document}